%% file: manuscript_stick-breaking-v39_rk_23.07.2019.tex
\documentclass[headings=standardclasses, 11pt]{scrartcl}

\usepackage{geometry}
\usepackage[utf8]{inputenc}
\usepackage[T1]{fontenc}
\usepackage[charter]{mathdesign}
\usepackage[onehalfspacing]{setspace}
\usepackage{microtype}
\usepackage{amsmath}
\usepackage{amsthm}
\usepackage{tabulary}
\usepackage{longtable}
\usepackage{booktabs} 
\newcommand{\ra}[1]{\renewcommand{\arraystretch}{#1}}
\usepackage{siunitx}
\usepackage{bm}
\usepackage[pdftex]{graphicx}
\usepackage{rotating}
\usepackage{pdfpages}
\usepackage{pdflscape}
\usepackage{color}
\usepackage{array}
\usepackage[font=bf]{caption}
\usepackage{pgfplots}
\usepackage{pgfplotstable}
\usepackage{float}
\usepackage{appendix}
\pgfplotsset{compat=1.9}
\usepackage{url}
\usepackage[section]{placeins}
\usepackage{natbib}
\usepackage{hyperref}
\hypersetup{
	colorlinks   = true,    
	urlcolor     = blue,    
	linkcolor    = blue,    
	citecolor    = blue      
}

\usepackage{nicefrac}

\begin{document}
\thispagestyle{empty}
\begin{spacing}{1.2}
\begin{flushleft}
\huge \textbf{Semi-Parametric Hierarchical Bayes Estimates of New Yorkers' Willingness to Pay for Features of Shared Automated Vehicle Services} \\
\vspace{\baselineskip}
\normalsize
23 July 2019 \\
\vspace{\baselineskip}
Rico Krueger \\
Research Centre for Integrated Transport Innovation, School of Civil and Environmental Engineering, UNSW Australia, Sydney NSW 2052, Australia \\
r.krueger@student.unsw.edu.au \\
ORCID ID: 0000-0002-5372-741X \\
\vspace{\baselineskip}
Taha H. Rashidi (corresponding author) \\
Research Centre for Integrated Transport Innovation, School of Civil and Environmental Engineering, UNSW Australia, Sydney NSW 2052, Australia\\
rashidi@unsw.edu.au\\
\vspace{\baselineskip}
Akshay Vij \\
Institute for Choice, University of South Australia\\
140 Arthur Street, North Sydney NSW 2060, Australia\\
vij.akshay@gmail.com \\
\end{flushleft}
\end{spacing}

\newpage
\thispagestyle{empty}
\section*{Abstract}

In this paper, we contrast parametric and semi-parametric representations of unobserved heterogeneity in hierarchical Bayesian multinomial logit models and leverage these methods to infer distributions of willingness to pay for features of shared automated vehicle (SAV) services. Specifically, we compare the multivariate normal (MVN), finite mixture of normals (F-MON) and Dirichlet process mixture of normals (DP-MON) mixing distributions. The latter promises to be particularly flexible in respect to the shapes it can assume and unlike other semi-parametric approaches does not require that its complexity is fixed prior to estimation. However, its properties relative to simpler mixing distributions are not well understood. In this paper, we evaluate the performance of the MVN, F-MON and DP-MON mixing distributions using simulated data and real data sourced from a stated choice study on preferences for SAV services in New York City. Our analysis shows that the DP-MON mixing distribution provides superior fit to the data and performs at least as well as the competing methods at out-of-sample prediction. The DP-MON mixing distribution also offers substantive behavioural insights into the adoption of SAVs. We find that preferences for in-vehicle travel time by SAV with ride-splitting are strongly polarised. Whereas one third of the sample is willing to pay between 10 and 80 USD/h to avoid sharing a vehicle with strangers, the remainder of the sample is either indifferent to ride-splitting or even desires it. Moreover, we estimate that new technologies such as vehicle automation and electrification are relatively unimportant to travellers. This suggests that travellers may primarily derive indirect, rather than immediate benefits from these new technologies through increases in operational efficiency and lower operating costs.
\\
\\
\textit{Keywords:} shared automated vehicles, willingness to pay, mixed logit, Dirichlet process, nonparametric methods.


\newpage
\pagenumbering{arabic}

\section{Introduction} \label{S:intro}

The representation of inter-individual taste heterogeneity is a key concern of discrete choice analysis, as information on the distribution of tastes is critical for demand forecasting, market segmentation and welfare analysis. In many empirical settings, the analyst cannot perfectly explain taste heterogeneity in terms of observed individual characteristics, and taste heterogeneity remains to a substantial extent random from the analyst's point-of-view. Mixed random utility models such as mixed logit or probit can accommodate any empirical random heterogeneity distribution by marginalising the discrete choice kernel over some mixing distribution, which describes the unobserved distribution of tastes in the sample \citep{mcfadden2000mixed, train2009discrete}. 

However, the ability of mixed random utility models to recover any true heterogeneity distribution is only predicated on an existence proof \citep[see][]{mcfadden2000mixed} and therefore, the analyst is required to select an appropriate mixing distribution in a given empirical setting. There are three principle ways in which unobserved taste heterogeneity can be incorporated into mixed random utility models \citep[for a review, see][]{vij2017random}: \emph{Parametric} mixing distributions such as the multivariate normal (MVN) distribution are described through a finite set of parameters, have well-defined functional forms, but are limited in the shapes they can assume. By contrast, \emph{nonparametric} mixing distributions such as the categorical distribution, which underlies the formulation of latent class models \citep[e.g.][]{greene2003latent}, are not described through a finite set of parameters and do not have well-defined functional forms. Accordingly, their complexity can adapt to the available information. Finally, \emph{semi-parametric} mixing distributions such as the finite mixture-of-normals (F-MON) distribution \citep[e.g.][]{rossi2012bayesian} aim to combine the benefits of the previous two approaches by convolving well-defined parametric kernels with flexible nonparametric mixing distributions. 

In the light of recent advances in technical computing soft- and hardware, Bayesian methods are re-emerging as a viable alternative to frequentist methods for the estimation of mixed random utility models \citep{bansal2019bayesian, ben2019foundations}. The key difference between Bayesian and frequentist procedures is that the Bayesian approach entails the specification of a full probability model for both the observed data and all unknown model parameters so that the posterior distribution of the unknown model parameters can be learnt by conditioning prior knowledge on observed data \citep{gelman2013bayesian}. Aside from several exceptions \citep{burda2008bayesian, daziano2013conditional, de2010bayesian, kim2004assessing, li2013bayesian}, Bayesian methods have been primarily used for the estimation of mixed random utility models with parametric mixing distributions \citep[e.g.][]{bansal2019bayesian, ben2019foundations, scarpa2008utility, train2005discrete, train2009discrete}. This is in spite of the fact that some Bayesian procedures such as Markov Chain Monte Carlo (MCMC) methods lend themselves well to the estimation of complex hierarchical models by facilitating the approximation of multi-dimensional integrals in models with many unknown parameters \citep[see][]{gelman2013bayesian}. 

The Dirichlet process mixture of normals (DP-MON) distribution \citep{antoniak1974mixtures, escobar1995bayesian} is a semi-parametric distribution, which is well-grounded in the hierarchical Bayesian modelling paradigm and promises to be particularly flexible in terms of the distributional shapes it can assume. The DP-MON distribution results from the convolution of a multivariate normal kernel with a Dirichlet process prior \citep{ferguson1973bayesian}, a flexible nonparametric mixing distribution which unlike the categorical distribution does not require that the number of mixture components is fixed prior to estimation. Rather, the complexity of the nonparametric mixing distribution is inferred from the evidence, and the number of mixture components is effectively treated as a model parameter. Accordingly, the DP-MON distribution can be viewed as an infinite-dimensional generalisation of the finite mixture-of-normals (F-MON) distribution. Notwithstanding that the DP-MON distribution has been incorporated into mixed random utility models to admit flexible representations of unobserved heterogeneity \citep[in particular][]{burda2008bayesian, de2010bayesian, li2013bayesian}, its performance relative to simpler mixing distributions such as the MVN or the F-MON distributions is not well understood. Importantly, the predictive ability of the DP-MON mixing distribution on external data is unknown, because existing studies solely rely on measures of in-sample fit or on visual inspections of the estimated heterogeneity distributions for model comparison \citep[see in particular][]{burda2008bayesian, de2010bayesian, li2013bayesian}. 

With recent advances in vehicle automation and electrification technologies and the advent of on-demand transportation services such as Uber, Lyft and Didi Chuxing, it has been envisioned that mobility on-demand services could be performed with automated and possibly electric vehicles \citep{burns2013vision, chen2016operations, fagnant2014travel, fagnant2016operations}. Such services are referred to as shared automated vehicle (SAV) services \citep{fagnant2014travel, fagnant2016operations}. In recent years, the development of methods for the strategic and operational control of SAV services has become an active field of research \citep[see e.g.][]{fagnant2014travel, fagnant2016operations, fagnant2018dynamic, hyland2018dynamic, levin2017congestion, levin2017general, liu2018framework, zhang2015exploring}, and it has been recognised that information on preferences and willingness to pay (WTP) for features of SAV services is critical for general transport planning as well as for the strategic and operational management of SAV services \citep[e.g.][]{krueger2016preferences, milakis2017policy}. However, in spite of several noteworthy efforts \citep{haboucha2017user, krueger2016preferences, lavieri2019modeling, liu2018framework, vij2018australian}, the literature on modelling preferences for SAV services remains at an incipient stage and lacks robust quantifications of WTP for SAV service attributes. Besides, the existing literature predominantly relies on parametric and nonparametric representations of unobserved heterogeneity \citep[see][]{haboucha2017user, krueger2016preferences, lavieri2019modeling, vij2018australian}. Thus, we hypothesise that research on modelling preferences for SAV services may benefit from the application of behavioural models with flexible representations of unobserved heterogeneity. 

Consequently, this paper has two interrelated objectives. In a first step, we seek to assess the practical implications of different flexible representations of unobserved heterogeneity on the ability of mixed logit to capture and predict preferences. To this end, we systematically evaluate the in- and out-of-sample performance of the MVN, F-MON and DP-MON mixing distributions for mixed logit in a simulation study and in a case study, which uses data from a stated choice survey \citep{bansal2018influence, liu2018framework} investigating preferences for SAV services in New York City. The case study also serves the second aim of the paper, which is to leverage the considered semi-parametric distributions to infer flexible estimates of WTP for different features of SAV services (out-of-vehicle travel time, in-vehicle travel time, vehicle automation and vehicle electrification).

The remainder of this paper is organised as follows: First, we present a review of the pertinent literature (Section \ref{S:review}). Next, we explain the modelling and estimation methodology (Section \ref{S:methodology}). Then, we present the simulation and case studies (Sections \ref{S:sim_study} and \ref{S:case_study}). Finally, we conclude (Section \ref{S:conclusion}). 

\section{Literature review} \label{S:review}

This section is divided into two subsections: In Section \ref{subS:sp}, we review semi-parametric methods for incorporating unobserved heterogeneity into mixed random utility models; for a review of parametric and nonparametric approaches, the reader is directed to \citet{vij2017random}. In Section \ref{subS:pref_mod}, we survey recent research investigating preferences for shared automated vehicle (SAV) services. 

\subsection{Semi-parametric mixing distributions} \label{subS:sp}

Semi-parametric mixing distributions produce smooth, continuous representations of unobserved heterogeneity and are extremely flexible in terms of the distributional shapes they can assume \citep[e.g.][]{vij2017random}. Therefore, the use of semi-parametric mixing distributions in mixed random utility models is compatible with the behavioural notion of preference continuity as opposed to the view that there are distinct subpopulations, each of which have characteristic preferences \citep[see][]{allenby1998marketing, wedel1999discrete}. 

The majority of semi-parametric approaches for incorporating unobserved taste heterogeneity into mixed random utility models have been proposed in the frequentist setting. One group of these approaches leverages flexible functionals such as Legendre polynomials \citep[see][]{fosgerau2007practical}, B-splines \citep[see][]{bastin2010estimating} and power series \citep[see][]{fosgerau2013easy} to accommodate unobserved heterogeneity in mixed logit models. These methods admit flexible representations of unobserved taste heterogeneity but require the analyst to configure the complexity of the employed functional prior to estimation. In addition, it is difficult to capture covariation between multiple random parameters, as each functional independently controls the distribution of exactly one parameter \citep[also see][for a discussion of the issue]{fosgerau2013easy}. Along similar lines, \citet{train2016mixed} proposes a semi-parametric mixed logit model in which an additional discrete mixing distribution is imposed on the parameters of a variety of flexible functionals such as step, spline or polynomial functions. The framework can flexibly recover differently-shaped, multivariate heterogeneity distributions but requires the analyst to select both the complexity of the discrete mixing distribution for the parameters of the functional as well as the functional itself prior to estimation \citep[see][]{bansal2018comparison, bansal2019flexible}. In practice, the approach can be computationally expensive, as it relies on bootstrapping for the calculation of standard errors \citep[see][]{train2016mixed}. 

Another group of semi-parametric approaches employs mixing distributions which are finite mixtures of parametric distributions. For example, \citet{bujosa2010combining}, \citet{fosgerau2009comparison} and \citet{keane2013comparing} leverage finite mixtures of independent normal distributions; \citet{greene2013revealing} employ finite mixtures of independent triangular distributions as a random taste parameter distribution. Several studies relax the diagonal restrictions on the covariance matrices of the multivariate normal components \citep{bansal2018minorization, bansal2019flexible, train2008algorithms}. In principle, the finite mixture of normals (F-MON) distributions with unrestricted multivariate normal components is an extremely flexible mixing distribution \citep[e.g.][]{rossi2012bayesian}. However, it requires the analyst to fix the number of mixture components prior to estimation. 

As highlighted in Section \ref{S:intro}, the Dirichlet process mixture of normals (DP-MON) distribution \citep{antoniak1974mixtures, escobar1995bayesian} represents an infinite-dimensional generalisation of the F-MON distribution and does not require that its complexity is fixed a priori. \citet{burda2008bayesian} use the DP-MON mixing distribution in combination with a multinomial logit kernel and apply the modelling approach to both synthetic data and real data sourced from a scanner panel survey. \citet{de2010bayesian} also use the DP-MON mixing distribution in combination with a multinomial logit kernel and validate the proposed model formulation on synthetic data. \citet{li2013bayesian} consider a multinomial probit kernel in combination with the DP-MON mixing distribution and test the modelling approach on synthetic data as well as on scanner panel data. Moreover, \citet{daziano2013conditional} applies a mixed logit model with a DP-MON mixing distribution to the analysis of stated choice data on vehicle purchase decisions in California. 

There are several noteworthy similarities between the before-mentioned studies employing the DP-MON mixing distribution in mixed random utility models. First, all existing studies use Bayesian methods for model estimation, as the DP-MON mixing distribution is well-grounded in the hierarchical Bayesian modelling paradigm due to its conditionally-conjugate structure. Second, extant studies contrast the performance of the DP-MON mixing distribution with the multivariate normal mixing distributions but not with the simpler F-MON mixing distribution. Third, extant studies rely on measures of in-sample fit and visual inspections of the estimated heterogeneity distributions for model comparison, while the predictive ability of the DP-MON mixing distribution on external data is not evaluated. 

To conclude, many of the existing semi-parametric approaches for incorporating unobserved taste heterogeneity into mixed random utility models are subject to limitations due to their immanent need for post-hoc model selection or their inability to capture dependencies between multiple taste parameters. The DP-MON mixing distribution addresses these shortcomings and promises to be particularly flexible in terms of the distributional shapes it can assume. However, its properties relative to simpler parametric and semi-parametric mixing distributions are not well understood. Thus, the current study seeks to systematically evaluate the in- and out-of-sample performance of the DP-MON mixing distribution in combination with a multinomial logit kernel in simulation experiments and a case study. 

\subsection{Preferences for shared automated vehicle services} \label{subS:pref_mod}

Shared automated vehicle (SAV) services refer to mobility on-demand (MOD) services which are performed with automated road vehicles \citep{fagnant2014travel, fagnant2016operations}. MOD represents an emerging paradigm for the provision of demand-responsive transportation services \citep{rayle2016just, shaheen2019shared, shaheen2016shared}: MOD services allow users to request rides in real-time or in advance from a pool of vehicles via mobile applications, which also handle payments. Unlike conventional on-demand transportation services such as taxis, MOD services use dynamic pricing regimes to manage different levels of supply and demand. Furthermore, MOD services can implement ride-splitting schemes, under which users who travel between similar origins and destinations are allocated to the same vehicle to travel together for parts of their trips. 

In many parts of the world, MOD services with conventional, human-driven vehicles represent an increasingly popular transport mode \citep[see][and the sources cited therein]{iqbal2019uber}. Examples of MOD brands providing private, non-shared rides include UberX, Lyft and Didi Express; examples of MOD brands providing pooled, shared rides include UberPool, Lyft Line and Didi Express Pool. With recent advances in vehicle automation and drive train electrification, it has been envisioned that MOD services could be performed with automated and possibly electric vehicles to increase the operational, economic and environmental efficiency of MOD services \citep[e.g.][]{burns2013vision, burns2013transforming, chen2016operations, fagnant2014travel, fagnant2016operations}.

Several studies employ stated choice methods to investigate preferences for SAV services. Drawing from a sample of 435 Australian consumers, \citet{krueger2016preferences} use a mixed logit model with triangular mixing distributions to analyse WTP for features of SAV services with and without ride-splitting. The study finds that the value of in-vehicle travel time for SAV services with ride-splitting is higher than for SAV services without ride-splitting. In addition, the results show that young adults and individuals with multimodal travel patterns may be comparatively more likely to adopt SAV services. Moreover, using data sourced from a nationally-representative sample of 3,985 Australian consumers, \citet{vij2018australian} develop a latent class ordered logit model to explain stated usage frequencies for hypothetical SAV service plans and estimate that consumers are on average willing to pay 0.28 AUD/km more to avoid ride-splitting and 0.17 AUD/km more for a door-to-door service. Relying on the same data as the current paper, \citet{liu2018framework} develop a multinomial logit choice model to obtain estimates of sensitivities to features of SAV services. The trained model is then fed into an integrated supply and demand framework for the control and optimisation of SAV service systems. Moreover, based on a sample of 721 individuals from Israel and the USA, \citet{haboucha2017user} implement a mixed logit model with univariate normal mixing distributions and estimate that 25\% of the considered sample would not use SAV services, even if such services were available free of charge. \citet{lavieri2019modeling} consider data from a convenience sample of 1,607 commuters living in the Dallas-Fort Worth metropolitan area and develop an integrated choice and latent variable model to analyse the willingness to use SAV services with ride-splitting. In the proposed model formulation, fixed travel time sensitivities are interacted with stochastic latent variables to accommodate both observed and unobserved individual heterogeneity. The study suggests that the disutility of ride-splitting may be greater for leisure trips than for commute trips and that travellers may derive greater disutility from the added travel time to pick up other passengers than from having to share a vehicle with strangers.

We also highlight several studies which investigate preferences for MOD services with conventional, human-driven vehicles. Relying on data sourced from the 2014/15 Puget Sound Regional Travel Study, \citet{dias2017behavioral} estimate a bivariate ordered logit model to explain stated usage frequencies of MOD and car-sharing services and find that young, well-educated, working individuals from high-income household in higher-density areas are relatively more likely to use these services. Drawing from stated preference data collected in Lisbon, Portugal, \citet{choudhury2018modelling} employ a nested logit model to estimate the value of travel time for smart mobility options including shared taxis. The study finds that travellers tend to prefer shared taxis for non-commute trips. Moreover, \citet{xie2019behavioral} present a holistic modelling framework capturing the sequential, yet inter-connected decision-process underlying the usage of MOD services within subscription-based Mobility-as-a-Service (MaaS) plans. The proposed modelling framework uses a nesting structure to connect multiple choice dimensions including subscription, service access as well as mode choice and accommodates random, parametric inter-individual heterogeneity. 

On the whole, research on modelling preferences for SAV services is still at an incipient stage. In terms of substantive behavioural insights, it is understood that level of service attributes such as fares, travel and waiting times are critical predictors of demand for SAV services, in particular when SAV services with and without ride-splitting compete with one another \citep{krueger2016preferences, lavieri2019modeling, vij2018australian}. Furthermore, some evidence suggests that some population subgroups such as young, well-educated urban residents tend to be comparatively more likely to use MOD services in general \citep{dias2017behavioral, krueger2016preferences}. In terms of the employed methodologies, it can be observed that several studies employ parametric representations of unobserved heterogeneity in sensitivities to features of SAV services \citep{haboucha2017user, krueger2016preferences, xie2019behavioral}, one study uses a nonparametric approach to the same end \citep{vij2018australian}, and another study considers stochastic latent variables to capture both observed and unobserved taste heterogeneity \citep{lavieri2019modeling}. An additional finding is that extant studies predominantly focus on explaining preferences, while the predictive performance of the employed modelling approaches is not evaluated. 


Consequently, another aim of the current study is to advance the literature on modelling preferences for SAV services. To this end, we apply mixed logit models with MVN, F-MON and DP-MON mixing distributions to data from a stated choice survey \citep{bansal2018influence, liu2018framework} on preferences for SAV services in New York City. We infer distributions of WTP for features of SAV services and compare the ability of the different heterogeneity representations to explain and predict preferences for SAV services. 

\section{Methodology} \label{S:methodology}


\subsection{Mixed logit model} \label{subS:mxl}

The mixed logit (MXL) model \citep{mcfadden2000mixed} can be established as follows: On choice occasion $t \in \{1, \ldots T_{n} \}$, decision-maker $n \in \{1, \ldots N \}$ derives utility $U_{ntj} = V(\boldsymbol{X}_{ntj}, \boldsymbol{\beta}_{n}) + \epsilon_{ntj}$ from alternative $j$ in set $C_{nt}$. $V()$ denotes the representative utility, $\boldsymbol{X}_{ntj}$ is a row-vector of covariates, $\boldsymbol{\beta}_{n}$ is a vector of individual-specific taste parameters, and $\epsilon_{ntj}$ is a stochastic disturbance. The assumption $\epsilon_{ntj} \sim \text{Gumbel}(0,1)$ leads to a multinomial logit (MNL) kernel such that the probability that decision-maker $n$ chooses alternative $j \in C_{nt}$ on choice occasion $t$ is 
\begin{equation}
P(y_{nt} = j \vert \boldsymbol{X}_{ntj}, \boldsymbol{\beta}_{n}) = \frac{\exp \left \{ V (\boldsymbol{X}_{ntj}, \boldsymbol{\beta}_{n}) \right \}}{\sum_{j' \in C_{nt}}\exp \left \{ V (\boldsymbol{X}_{ntj'}, \boldsymbol{\beta}_{n}) \right \}},
\end{equation} 
where $y_{nt} \in C_{nt}$ captures the observed choice. The choice probability can be iterated over choice occasions to obtain the probability of observing a decision-maker's sequence of choices $\boldsymbol{y}_{n}$:
\begin{equation}
P(\boldsymbol{y}_{n} \vert \boldsymbol{X}_{n},  \boldsymbol{\beta}_{n}) = \prod_{t = 1}^{T_{n}} P(y_{nt} = j \vert \boldsymbol{X}_{nt},  \boldsymbol{\beta}_{n}).
\end{equation}

The individual-specific taste parameters $\boldsymbol{\beta}_{1:N}$ can follow any probability distribution, but the multivariate normal (MVN) distribution is the most commonly assumed mixing distribution. In this case, $\boldsymbol{\beta}_{n} \sim \text{N}(\boldsymbol{\zeta}, \boldsymbol{\Omega})$ for $n = 1, \dots, N$, where $\boldsymbol{\zeta}$ is a mean vector and $\boldsymbol{\Omega}$ is a covariance matrix. Under a fully Bayesian setup, $\boldsymbol{\zeta}$ and $\boldsymbol{\Omega}$ are also considered to be random parameters and are thus given priors. We use a normal prior for mean vector $\boldsymbol{\zeta}$ and employ Huang's half-t prior \citep{huang2013Simple} for covariance matrix $\boldsymbol{\Omega}$, as this prior specification exhibits superior  noninformativity properties compared to alternative prior specifications \citep{akinc2018Bayesian, huang2013Simple}.

Stated succinctly, the generative process of a fully Bayesian MXL model with an unrestricted multivariate normal mixing distribution is:
\begin{align}
& \boldsymbol{\zeta} \vert \boldsymbol{\mu}_{0},\boldsymbol{\Sigma}_{0} \sim \text{N}(\boldsymbol{\mu}_{0},\boldsymbol{\Sigma}_{0}) \\
& a_{r} \lvert A_{r} \sim \text{Gamma}\left( \frac{1}{2}, \frac{1}{A_{r}^{2}} \right), r = 1,\dots,R,  \label{eq:mvn_gamma_a} \\
& \boldsymbol{\Omega} \vert \nu, \boldsymbol{a}\sim \text{IW}\left(\nu + R - 1, 2\nu \text{diag}(\boldsymbol{a}) \right),  \quad \boldsymbol{a} = \begin{bmatrix} a_{1} & \dots & a_{R} \end{bmatrix}^{\top} \label{eq:mvn_iv_Omega} \\
& \boldsymbol{\beta}_{n} \vert \boldsymbol{\zeta}, \boldsymbol{\Omega} \sim \text{N}(\boldsymbol{\zeta}, \boldsymbol{\Omega}), n = 1,\dots,N, \\
& y_{nt} \vert \boldsymbol{\beta}_{n}, \boldsymbol{X}_{nt} \sim \text{MNL}(\boldsymbol{\beta}_{n}, \boldsymbol{X}_{nt}), n = 1,\dots,N,  \ t = 1,\dots,T_{n}.
\end{align} 
Here, (\ref{eq:mvn_gamma_a}) and (\ref{eq:mvn_iv_Omega})  induce Huang's half-t prior \citep{huang2013Simple}, and $r \in \{1, \ldots, R \}$ indexes the random parameters. Moreover, $\{ \boldsymbol{\mu}_{0}, \boldsymbol{\Sigma}_{0}, \nu, A_{1:R} \}$ are known hyper-parameters, and $\boldsymbol{\theta} = \{ \boldsymbol{\zeta}, \boldsymbol{\Omega},  \boldsymbol{a}, \boldsymbol{\beta}_{1:N}\}$ is a collection of model parameters whose posterior distribution we wish to estimate. 
The generative process implies the following joint distribution of data and model parameters:
\begin{equation}
P (\boldsymbol{y}_{1:N}, \boldsymbol{\theta}) = 
\left ( \prod_{n=1}^{N}  P(\boldsymbol{y}_{n} \vert \boldsymbol{X}_{n}, \boldsymbol{\beta}_{n}) \right )
\left ( \prod_{n=1}^{N} P(\boldsymbol{\beta}_{n} \vert \boldsymbol{\zeta}, \boldsymbol{\Omega}) \right )
P(\boldsymbol{\zeta} \vert \boldsymbol{\mu}_{0},\boldsymbol{\Sigma}_{0})
P(\boldsymbol{\Omega} \vert \omega, \boldsymbol{B})
\left ( \prod_{r=1}^{R} P(a_{r} \lvert  s,  u_{r}) \right )
\end{equation}
where 
$\omega = \nu + R - 1$, 
$\boldsymbol{B} = 2\nu \text{diag}(\boldsymbol{a})$, 
$s = \frac{1}{2}$ and
$u_{r} = A_{r}^{-2}$.
By Bayes' rule, the posterior distribution of interest is then given by
\begin{equation} \label{eq:post}
P(\boldsymbol{\theta} \vert \boldsymbol{y}_{1:N}) 
= \frac{P (\boldsymbol{y}_{1:N}, \boldsymbol{\theta})}{\int P (\boldsymbol{y}_{1:N}, \boldsymbol{\theta}) d \boldsymbol{\theta}}
\propto P (\boldsymbol{y}_{1:N}, \boldsymbol{\theta}).
\end{equation}

\subsection{Semi-parametric representations of unobserved heterogeneity} \label{subS:snp}

\subsubsection{Finite mixture of normals distribution}

As an alternative to the restrictive multivariate normal mixing distribution, the more expressive finite mixture of normals (F-MON) mixing distribution can be employed \citep[e.g.][]{rossi2012bayesian}. This mixing distribution assumes that decision-makers are distributed over $K$ mixture components indexed by $k \in \{1, \ldots K \}$. Each mixture component is a multivariate normal distribution, which describes the distribution of tastes in the mixture component and is characterised by its own, component-specific mean vector $\boldsymbol{\zeta}_{k}$ and covariance matrix $\boldsymbol{\Omega}_{k}$. A decision-maker's assignment $q_{n} \in \{1,\ldots, K\}$ to one of the $K$ mixture components is drawn from a categorical distribution with parameter $\boldsymbol{\pi}$, a probability simplex sampled from a symmetric Dirichlet prior with parameter $\alpha$. Consequently, the individual-specific tastes $\boldsymbol{\beta}_{n}$ are generated conditional on component assignment $q_{n}$, i.e. $\boldsymbol{\beta}_{n} \sim \text{N}(\boldsymbol{\zeta}_{q_{n}}, \boldsymbol{\Omega}_{q_{n}}), n = 1,\dots, N$. 

Stated succinctly, the generative process of the individual-specific taste parameters $\boldsymbol{\beta}_{1:N}$ under the F-MON mixing distribution is as follows:
\begin{align}
& \boldsymbol{\zeta}_{k} \vert \boldsymbol{\mu}_{0},\boldsymbol{\Sigma}_{0} \sim \text{N}(\boldsymbol{\mu}_{0},\boldsymbol{\Sigma}_{0}) \\
& a_{kr} \vert A_{r} \sim \text{Gamma}\left( \frac{1}{2}, \frac{1}{A_{r}^{2}} \right), k = 1,\dots, K, r = 1,\dots, R, \label{eq:fmon_gamma_a} \\
& \boldsymbol{\Omega}_{k} \vert \nu, \boldsymbol{a}_{k} \sim \text{IW}\left(\nu + R - 1, 2\nu \text{diag}(\boldsymbol{a}_{k}) \right), \boldsymbol{a}_{k} = \begin{bmatrix} a_{k,1} & \dots & a_{k,R} \end{bmatrix}^{\top}, k = 1,\dots, K, \label{eq:fmon_iv_Omega} \\
& \boldsymbol{\pi} \vert \boldsymbol{\alpha} \sim \text{Dirichlet}(\alpha) \\
& q_{n} \vert \boldsymbol{\pi} \sim \text{Categorical}(\boldsymbol{\pi}) , n = 1,\dots,N, \\
& \boldsymbol{\beta}_{n} \vert \boldsymbol{\zeta}_{q_{n}}, \boldsymbol{\Omega}_{q_{n}} \sim \text{N}(\boldsymbol{\zeta}_{q_{n}}, \boldsymbol{\Omega}_{q_{n}}), n = 1,\dots,N.
\end{align} 
Again, we use a normal prior for mean vector $\boldsymbol{\zeta}_{k}$ and Huang's half-t prior \citep{huang2013Simple} for covariance matrix $\boldsymbol{\Omega}_{k}$ (see expressions \ref{eq:fmon_gamma_a} and \ref{eq:fmon_iv_Omega}).

The number of mixture components $K$ of the F-MON mixing distribution represents an exogenous model parameter, which can be determined based on a consideration of post-hoc model selection criteria which are applied to a candidate set of models with varying numbers of mixture components. However, in spite of this flexibility, extant studies predominantly employ model specifications with two mixture components due to considerations of parsimony and computational tractability \citep[see e.g.][]{bansal2018minorization, bansal2019flexible, fosgerau2009comparison, keane2013comparing, train2008algorithms}. Thus, we assume a fixed computational budget in the subsequent applications and also only consider F-MON mixing distributions with exactly two mixture components (i.e. $K = 2$). Henceforth, we refer to this mixing distribution as 2-F-MON distribution. 

\subsubsection{Dirichlet process mixture of normals distribution}

The Dirichlet process mixture of normals (DP-MON) distribution \citep{antoniak1974mixtures, escobar1995bayesian} results from the convolution of a multivariate normal kernel with a Dirichlet process prior \citep{ferguson1973bayesian}, a flexible nonparametric mixing distribution which, unlike the categorical distribution, does not require that the number of mixture components is fixed prior to estimation. 

\begin{figure}[H]
\centering
\includegraphics[width = \textwidth]{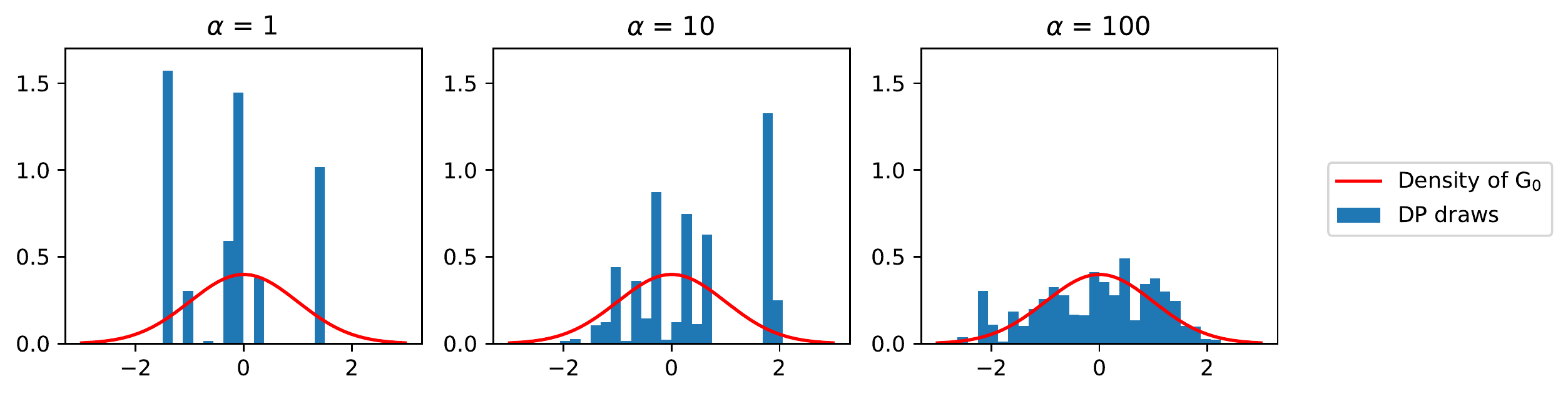}
\caption{Distributions of 1,000 draws from realisations of Dirichlet processes with baseline $\text{N}(0,1)$ and different values of $\alpha$} \label{f_dp}
\end{figure}

To be precise, the Dirichlet process \citep{ferguson1973bayesian} is a stochastic process, whose realisations $\text{G} \sim \text{DP}(\alpha, G_{0})$ are probability distributions on a space $\Theta$ (such as the $n$-dimensional real space). It is parameterised by a concentration parameter $\alpha$ and a base measure $G_{0}$, which itself is a probability distribution on $\Theta$. $\text{G}_{0}$ is an initial guess about $\text{G}$ and $\alpha$ controls the proximity of $\text{G}_{0}$ and $\text{G}$. Figure \ref{f_dp} illustrates the behaviour of a Dirichlet process with a standard normal base measure for different values of $\alpha$.

From Figure \ref{f_dp}, it can also be seen that the Dirichlet process exhibits two important properties, which qualify it as a nonparametric prior in clustering and segmentation models. First, realisations from the Dirichlet process are discrete and second, repeated samples from a realisation $\text{G}$ from the Dirichlet process are clustered with non-zero probability. \citet{ferguson1973bayesian} provides formal proofs of both properties. However, the properties of the Dirichlet process can also be illustrated in more intuitive ways through its constructive representations, namely the Blackwell-MacQueen urn scheme \citep{blackwell1973ferguson}, the Chinese Restaurant process \citep{aldous1985exchangeability} and the stick-breaking process \citep{sethuraman1994constructive}. In this study, we focus on the stick-breaking process representation, as it facilitates estimation \citep[see][]{gelman2013bayesian, ishwaran2001gibbs}

\begin{figure}[H]
\centering
\includegraphics[width = 0.85 \textwidth]{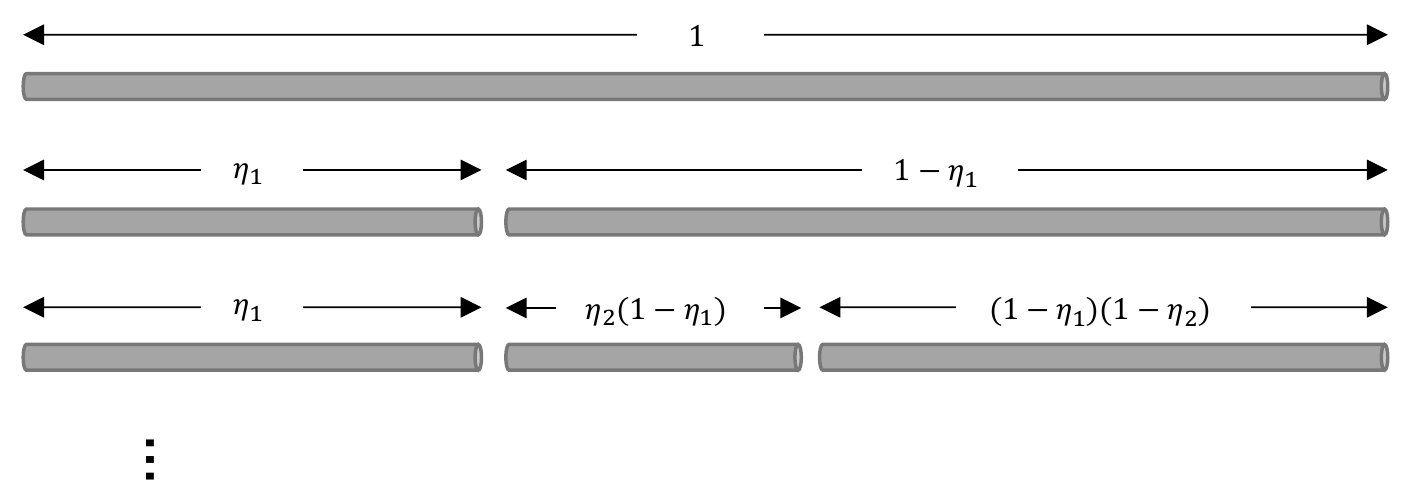}
\caption{Illustration of the stick-breaking construction of the Dirichlet process} \label{figure_stick-breaking}
\end{figure}

The stick-breaking process constructs a realisation from a Dirichlet process $\text{G} \sim \text{DP}(\alpha, \text{G}_{0})$ as a discrete mixture of point masses, whereby the component weights are factorisations of Beta random variables, i.e.
\begin{equation}
\text{G} = \sum_{k=1}^{\infty} \pi_{k} \delta_{\theta_{k}}
\end{equation}
with 
\begin{equation} \label{e_stick-breaking}
\eta_{k} \sim \text{Beta}(1,\alpha),
\quad \pi_{k} = \eta_{k} \prod_{l=1}^{k-1}(1 - \eta_{l}),
\quad \theta_{k} \sim \text{G}_{0},
\quad k = 1,\ldots,\infty,
\end{equation}
where $\pi_{k} \in (0,1)$, $k = 1,\ldots,\infty,$ are probability weights with $\sum_{k=1}^{\infty} \pi_{k} = 1$; $\delta_{\theta_{k}}$ are the associated point masses centred at $\theta_{k}$, which in turn are realisations from $\text{G}_{0}$. The stick-breaking construction can be illustrated as follows (see Figure \ref{figure_stick-breaking}): Starting with stick of unit length, we break the stick at $\eta_{1} \sim \text{Beta}(1,\alpha)$ and assign $\pi_{1} = \eta_{1}$ to the broken-off piece. Next, we sample $\eta_{2} \sim \text{Beta}(1,\alpha)$ and break a piece of length $\pi_{2} = \eta_{2} (1 - \eta_{1})$ from the remainder of the stick. Subsequently, we continue to break off pieces of lengths $\pi_{3}, \ldots , \pi_{k}$  of the remainder of the stick. 

The truncated stick-breaking construction \citep{ishwaran2001gibbs} is a finite-dimensional approximation of the infinite-dimensional stick-breaking construction. Under this truncation approximation, the sequential stick-breaking process is terminated after $K$ steps. To assure that $\sum_{k=1}^{K} \pi_{k} = 1$, the final Beta random variable is degenerate with $\eta_{K} = 1$ so that $\pi_{K} = 1 - \sum_{k=1}^{K - 1} \pi_{k}$. At first glance, the use of a truncation approximation appears to defeat the purpose of nonparametric mixing distribution with an endogenous clustering mechanism, as we are essentially defining a finite mixture model of dimension $K$. However, the truncated stick-breaking construction induces a shrinkage on the number of effectively populated mixture components, while maintaining the computational advantages of a finite mixture model \citep{gelman2013bayesian}. In fact, the residual probability $\pi_{K}$ is negligibly small for reasonably large $K$ and most $\alpha$ values that are encountered in practice \citep{ishwaran2000markov, ohlssen2007flexible}. Subsequently, we follow \citet{de2010bayesian} and set $K = 100$ to assure a close approximation of the Dirichlet process.

With the DP-MON mixing distribution, the generative process of the individuals-specific taste parameters $\boldsymbol{\beta}_{1:N}$ is nearly the same as with the F-MON mixing distribution. The key difference is that the model of the component membership probabilities $\boldsymbol{\pi}$ is modified to accommodate the truncated stick-breaking process approximation of the Dirichlet process prior on the number of mixture components. The base measure $\mbox{G}_{0}$ must also be specified. In general, it is desirable that the $\mbox{G}_{0}$ is conjugate to the kernel to facilitate model inference \citep{gelman2013bayesian}. Thus, the normal distribution is an obvious choice. \citet{gelman2013bayesian} advise against the use of diffuse base measures, as a high variance of the base measure may penalise the addition of new mixture components and may thus limit the flexibility of the mixing distribution. Consequently, we set $\text{G}_{0} = \text{N}(\boldsymbol{0}, \boldsymbol{I})$ in the subsequent applications. Moreover, we follow \citet{ishwaran2002approximate} and treat the concentration parameter $\alpha$ as unknown; we let $\alpha \sim \text{Gamma}(2,2)$, i.e. the prior density of $\alpha$ is Gamma with shape 2 and scale 2. 

Stated succinctly, the generative process of the resulting MXL model is:
\begin{align}
& \boldsymbol{\zeta}_{k} \vert \boldsymbol{\mu}_{0},\boldsymbol{\Sigma}_{0} \sim \text{N}(\boldsymbol{\mu}_{0},\boldsymbol{\Sigma}_{0}) \\
& a_{kr} \vert A_{r} \sim \text{Gamma}\left( \frac{1}{2}, \frac{1}{A_{r}^{2}} \right), k = 1,\dots, K, r = 1,\dots, R, \\
& \boldsymbol{\Omega}_{k} \vert \nu, \boldsymbol{a}_{k} \sim \text{IW}\left(\nu + R - 1, 2\nu \text{diag}(\boldsymbol{a}_{k}) \right), \boldsymbol{a}_{k} = \begin{bmatrix} a_{k,1} & \dots & a_{k,R} \end{bmatrix}^{\top}, k = 1,\dots, K, \\
& \alpha \vert 2,2 \sim \text{Gamma}(2,2) \\
& \eta_{k} \vert \alpha \sim \text{Beta}(1, \alpha), k = 1,\dots, K - 1, \eta_{K} = 1, \\
& \pi_{k} = \eta_{k} \prod_{l=1}^{k-1}(1 - \eta_{l}), k = 1,\dots, K, \\
& q_{n} \vert \boldsymbol{\pi} \sim \text{Categorical}(\boldsymbol{\pi}) , n = 1,\dots,N, \\
& \boldsymbol{\beta}_{n} \vert \boldsymbol{\zeta}_{q_{n}}, \boldsymbol{\Omega}_{q_{n}} \sim \text{N}(\boldsymbol{\zeta}_{q_{n}}, \boldsymbol{\Omega}_{q_{n}}), n = 1,\dots,N, \\
& y_{nt} \vert \boldsymbol{\beta}_{n}, \boldsymbol{X}_{nt} \sim \text{MNL}(\boldsymbol{\alpha}, \boldsymbol{\beta}_{n}, \boldsymbol{X}_{nt}), n = 1,\dots,N,  \ t = 1,\dots,T_{n}.
\end{align} 

Finally, we point out differences between the model presented above and extant instances of DP-MON-MXL models. \citet{de2010bayesian} consider an almost identical model with the key difference being that the concentration parameter $\alpha$ is fixed rather than estimated. \citet{burda2008bayesian} also do not estimate $\alpha$ and exploit the Chinese Restaurant process instead of the truncated stick-breaking process approximation to construct the Dirichlet process prior. \citet{daziano2013conditional} does not provide details on the construction of the Dirichlet process prior and does not infer $\alpha$. Moreover, none of the extant instances of  DP-MON-MXL models use Huang's half-t prior \citep{huang2013Simple} for covariance matrix $\boldsymbol{\Omega}_{k}$.

\subsection{Bayesian estimation via Markov chain Monte Carlo methods} \label{subS:estimation}

For posterior inference in the fully Bayesian MXL models with MVN-, F-MON- and DP-MON-MXL models mixing distributions, we employ Markov chain Monte Carlo (MCMC) methods, whose underlying idea is to approximate a difficult-to-compute posterior distribution through samples from a Markov chain whose stationary distribution is the posterior distribution of interest \citep[see][for a general treatment]{robert2004monte}. In the present application, such Markov chains can be easily constructed due to the conditionally-conjugate structure of the considered hierarchical models. For most model parameters, posterior draws can be generated via direct sampling from the conditional distributions. Only updates for the individual-specific taste parameters $\boldsymbol{\beta}_{1:N}$ need to be generated with the help of the random-walk Metropolis algorithm, because the multinomial logit kernel does not have a general conjugate prior. In Appendix \ref{A:estimation}, we provide the MCMC sampling schemes for the models considered in this study. Fixed utility parameters can be easily accommodated through the inclusion of an additional Metropolis step \citep[see][]{train2009discrete}. In addition, it is possible for a model to contain a combination of parametrically and semi-parametrically distributed individual-specific parameters, which can then be updated in a joint Metropolis step.

Bayesian inference in discrete mixture models such as the F-MON- and DP-MON-MXL models is complicated by the so-called label-switching issue, which arises because the labels of the multivariate normal mixture components can be permuted without any effect on the likelihood value \citep[see][]{gelman2013bayesian, rossi2012bayesian}: As a consequence, the labels of the multivariate normal mixture components can switch from one MCMC iteration to another, in particular when the components are not well separated, and the posterior densities of the component-specific parameters $\pi_{k}$, $\boldsymbol{\zeta}_{k}$ and $\boldsymbol{\Omega}_{k}$ are not identifiable. However, regardless of potential label-switching, the joint mixture density $f(y \vert \boldsymbol{\pi}, \boldsymbol{\zeta}_{1:K}, \boldsymbol{\Omega}_{1:K}) = \sum_{k = 1}^{K} \pi_{k} f(y \vert \boldsymbol{\zeta}_{k}, \boldsymbol{\Omega})$ remains identified at all times. Hence, it is possible to obtain the posterior distribution of the joint mixture density by evaluating $f(y \vert \boldsymbol{\pi}, \boldsymbol{\zeta}_{1:K}, \boldsymbol{\Omega}_{1:K})$ along a dense grid of $y$ values at the posterior draws of $\boldsymbol{\pi}$, $\boldsymbol{\zeta}_{1:K}$ and $\boldsymbol{\Omega}_{1:K}$. Evidently, the label-switching issue also complicates convergence assessment \citep{gelman2013bayesian}, which is in general an open problem in Bayesian estimation \citep[see][]{depraetere2017comparison, rossi2012bayesian}. As the posterior distributions of the component-specific parameters are not identifiable, convergence diagnostics cannot be calculated based on the posterior draws of these parameters. One way to circumvent this issue is to assess convergence based on derived quantities, which are not affected by label-switching \citep{gelman2013bayesian}. 

For more information about Bayesian estimation of mixed random utility models, the reader is directed to the literature \citep{ben2019foundations, rossi2012bayesian, train2009discrete}. We also highlight that the conjunction of hierarchical models and Bayesian estimation is often referred to as hierarchical Bayesian approach \citep[e.g.][]{ben2019foundations, gelman2013bayesian, rossi2012bayesian, train2009discrete}. 

\section{Simulation study} \label{S:sim_study}


\subsection{Data and experimental setup} \label{S:sim_study:data}

For the simulation study, we rely on synthetic choice data, which we generate as follows: Choice tasks include five unlabelled alternatives, which are characterised by two attributes. Decision-makers are assumed to be utility-maximisers and to evaluate alternatives based on the utility specification $U_{ntj} = X_{ntj,1} \beta_{n,1} + X_{ntj,2} \beta_{n,2} + \epsilon_{ntj}$, where $\{ X_{ntj,1}, X_{ntj,2} \}$ are alternative-specific attributes and $\{ \beta_{n,1}, \beta_{n,2} \}$ are the corresponding individual-specific taste parameters. $\epsilon_{ntj}$ is a stochastic disturbance sampled from $\text{Gumbel}(0,1)$. $n \in \{1, \ldots, N\}$ indexes decision-makers, $t \in \{1, \ldots, T\}$ indexes choice tasks and $j \in \{1, \ldots, J\}$ indexes alternatives. 

For the generation of $\{ \beta_{n,1}, \beta_{n,2} \}$, we consider the same two scenarios as \citet{burda2008bayesian}. Both scenarios involve taking draws from a skew-normal-logistic distribution \citep[e.g.][]{nadarajah2003skewed} with density $f(x \vert \mu, \sigma, \lambda) = 2 \phi(x \vert \mu, \sigma) G \left ( \lambda (x - \mu) \right)$, where $\phi = \frac{1}{\sqrt{2 \pi \sigma^2}}{ \exp \left(- \frac{(x - \mu)^{2}}{2 \sigma^2}  \right)}$ denotes the probability distribution function of a normally distributed random variable and $G(y) = \frac{1}{1 + \exp (-y)}$ is the cumulative distribution function of a logistically distributed random variable. In that vein, we let $\text{SNL}(\mu, \sigma, \lambda)$ denote a skew-normal-logistic distribution with mean $\mu$, scale $\sigma$ and shape $\lambda$. In scenario 1, the distribution of the true taste parameters is skewed: We have $\beta_{n,r} \sim \text{SNL}(0, 1, 50)$ for $r = 1, 2$. 
In scenario 2, the distribution of the true taste parameters is skewed and multi-modal: We have 
$\beta_{n,1} \sim \text{SNL}(1, 1, 40)$, $\beta_{n,2} \sim \text{SNL}(-2, 1, 80)$ for 25\% of the decision-makers, 
$\beta_{n,1} \sim \text{SNL}(-2, 1, 70)$, $\beta_{n,2} \sim \text{SNL}(-2, 1, 70)$ for another 25\% of the decision-makers and
$\beta_{n,1} \sim \text{SNL}(1, 1, -50)$, $\beta_{n,2} \sim \text{SNL}(1, 1, -50)$ for the remaining 50\% of decision-makers.
In both scenarios, the alternative-specific attributes $\{ X_{ntj,1}, X_{ntj,2} \}$ are drawn from $\text{Uniform}(-5,5)$, which implies an error rate of approximately 25\%, i.e. in 25\% of the cases, decision-makers deviate from the deterministically-best alternative due to the stochastic utility component. We set $N = 1000$ and allow $T$ to take a value in $\{8, 16\}$. For each scenario and value of $T$, we consider 20 replications, whereby the data for each replication are generated using a different random seed.

Each of the synthetic datasets is used to estimate MVN-, 2-F-MON- and DP-MON-MXL models. We implement the MCMC methods described in Appendix \ref{A:estimation} by writing our own Python code. The MCMC algorithms are executed with two parallel Markov chains and 100,000 iterations for each chain, whereby the first 50,000 iterations of each chain are discarded for burn-in. After burn-in, only every tenth draw is retained to reduce the amount of autocorrelation in the chains. The simulation experiments are conducted on the Katana high performance computing cluster at the Faculty of Science, UNSW Australia.

\subsection{Performance assessment}

We evaluate the performance of the modelling approaches in terms of their out-of-sample predictive accuracy, as is common practice in the context of Bayesian analysis of discrete choice models \citep[see][]{bansal2019bayesian, braun2010variational, depraetere2017comparison, tan2017stochastic}. There are three key advantages to considering out-of-sample predictive accuracy for model checking and criticism. First, out-of-sample predictive accuracy serves as a self-consistency check for a model by revealing discrepancies between observed data and predictions generated from the trained model \citep{gelman2013bayesian}. Second, predictive accuracy allows for a succinct summary of model performance, when the number of model parameters is large, and third, predictive accuracy accounts for the uncertainty in the estimates, as the predicted quantity is integrated over the estimated posterior distribution of the model parameters \citep{depraetere2017comparison}.

To evaluate the out-of-sample predictive accuracy of the modelling approaches, we compute the total variation distance \citep[TVD;][]{braun2010variational} between the true and the estimated predictive choice distributions for a validation sample, which we generate along with each training sample. Each validation sample is based on the same data generating process as its respective training sample, whereby the number of decision-makers is set to 25 and the number of observations per decision-maker is set to one. The true predictive choice distribution for a choice set $C_{nt}$ with attributes $\boldsymbol{X}_{nt}^{*}$ from the validation sample is given by
\begin{equation}
P_{\text{true}}(y_{nt}^{*} \vert \boldsymbol{X}_{nt}^{*}) = 
\int P(y_{nt}^{*} = j \vert \boldsymbol{X}_{nt}^{*}, \boldsymbol{\beta}) f(\boldsymbol{\beta}) d \boldsymbol{\beta}.
\end{equation}
This integration is not tractable and is therefore simulated using 10,000 pseudo-random draws from the true heterogeneity distribution $f(\boldsymbol{\beta})$. For the MVN-MXL model, the corresponding estimated predictive choice distribution is 
\begin{equation} \label{e:ppd_mvn}
\hat{P} (y_{nt}^{*} \vert \boldsymbol{X}_{nt}^{*}, \boldsymbol{y}) = 
\int \int
\left ( \int P(y_{nt}^{*} \vert \boldsymbol{X}_{nt}^{*}, \boldsymbol{\beta}) f(\boldsymbol{\beta} \vert \boldsymbol{\zeta}, \boldsymbol{\Omega}) d \boldsymbol{\beta} \right ) p(\boldsymbol{\zeta}, \boldsymbol{\Omega} \vert \boldsymbol{y}) d \boldsymbol{\zeta} d \boldsymbol{\Omega};
\end{equation}
for the 2-F-MON- and DP-MON-MXL models, the corresponding estimated predictive choice distribution is 
\begin{equation} \label{e:ppd_mon}
\begin{split}
\hat{P} (y_{nt}^{*} \vert \boldsymbol{X}_{nt}^{*}, \boldsymbol{y}) = &
\int \int \int
\left ( \int P(y_{nt}^{*} \vert \boldsymbol{X}_{nt}^{*}, \boldsymbol{\beta}) f(\boldsymbol{\beta} \vert \boldsymbol{\pi}, \boldsymbol{\zeta}_{1:K}, \boldsymbol{\Omega}_{1:K}) d \boldsymbol{\beta} \right ) \\
& p(\boldsymbol{\pi}, \boldsymbol{\zeta}_{1:K}, \boldsymbol{\Omega}_{1:K} \vert \boldsymbol{y}) d \boldsymbol{\pi} d \boldsymbol{\zeta}_{1:K} d \boldsymbol{\Omega}_{1:K}.
\end{split}
\end{equation}
In both cases, the estimated posterior predictive distributions are computed via Monte Carlo integration, whereby $p(\boldsymbol{\zeta}, \boldsymbol{\Omega} \vert \boldsymbol{y})$ and $p(\boldsymbol{\pi}, \boldsymbol{\zeta}_{1:K}, \boldsymbol{\Omega}_{1:K} \vert \boldsymbol{y})$ are given by the empirical distribution of the posterior draws; for all models, we use 2,000 i.i.d simulation draws for $\boldsymbol{\beta}$. TVD is then given by 
\begin{equation}
\text{TVD} = \frac{1}{2} \sum_{j \in C_{nt}} \left | 
P_{\text{true}}(y_{nt}^{*} = j \vert \boldsymbol{X}_{nt}^{*})  - 
\hat{P} (y_{nt}^{*} = j \vert \boldsymbol{X}_{nt}^{*}, \boldsymbol{y})  \right |.
\end{equation}
For succinctness, we calculate averages across decision-makers and choice sets. Lower values of TVD indicate superior predictive accuracy. 

\subsection{Results}

Tables \ref{table_results_S1} and \ref{table_results_S2} enumerate the simulation results for scenarios 1 and 2, respectively. In each table, we report the means and the standard errors of the estimation times and TVD across 20 replications for $N = 1000$ and $T \in \{8, 16\}$. We observe that in both scenarios, DP-MON-MXL is the most accurate method followed by 2-F-MON-MXL. Moreover, the relative differences in prediction accuracy between the different methods are more pronounced in scenario 2 (multi-modal, skewed tastes) than in scenario 1 (skewed tastes) . We further observe that in scenario 2, the prediction accuracy of DP-MON-MXL benefits from a greater number of choice occasions per decision-maker, whereas the prediction accuracy of the other methods deteriorates, as $T$ increases. Furthermore, we observe that DP-MON-MXL is the slowest of all methods, but nonetheless estimation times remain manageable for sample sizes that are typically encountered in stated preference studies. 

\begin{figure}[H]
\centering
\includegraphics[width = \textwidth]{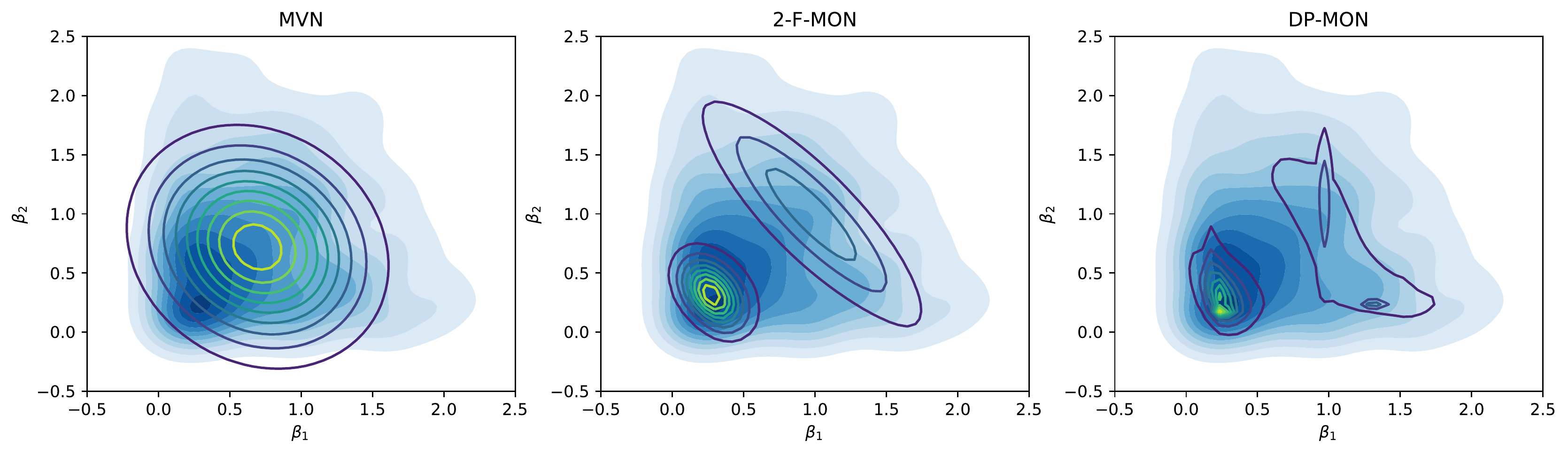}
\caption{Kernel density estimate of true heterogeneity distribution (shaded area) and estimated posterior predictive heterogeneity distributions (contour lines) for one of the replications of scenario 1 (skewed tastes) and T = 8} \label{f_sim1}
\end{figure}

\begin{figure}[H]
\centering
\includegraphics[width = \textwidth]{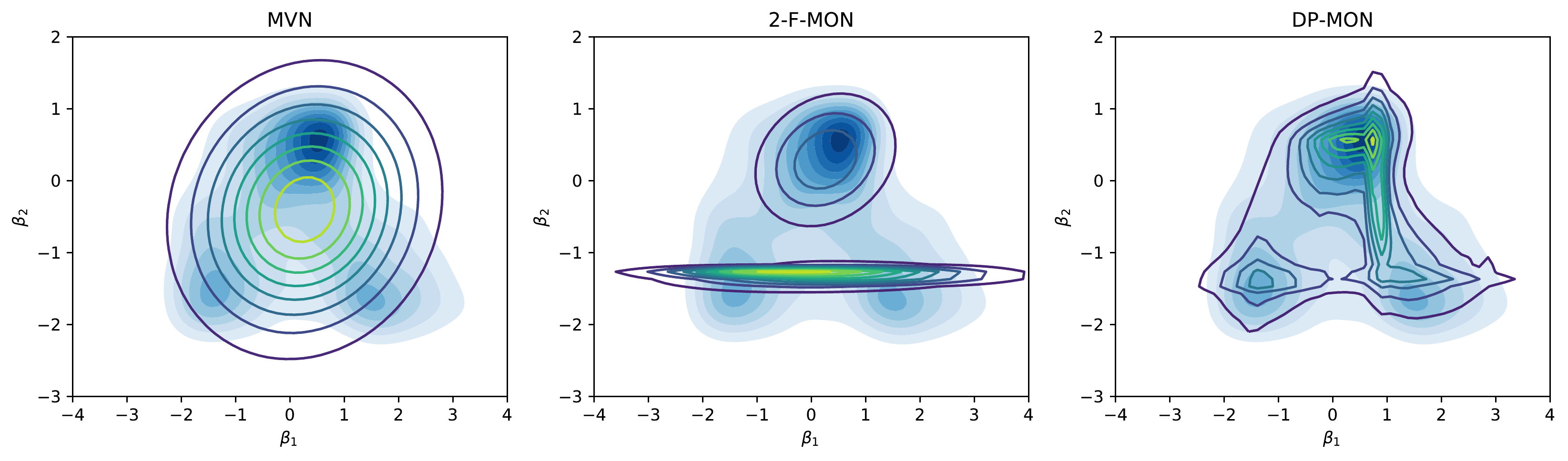}
\caption{Kernel density estimate of true heterogeneity distribution (shaded area) and estimated posterior predictive heterogeneity distributions (contour lines) for one of the replications of scenario 2 (multi-modal, skewed tastes) and T = 8} \label{f_sim2}
\end{figure}

The differences in predictive accuracy are also reflected in qualitative differences in the heterogeneity representations produced by each of the models. Figures \ref{f_sim1} and \ref{f_sim2} show kernel density estimates of the true heterogeneity distributions and contour plots of the estimated posterior predictive heterogeneity distributions for each of the models for one of the replications of scenarios 1 and 2, respectively, and for $T = 8$. It can be seen that in scenario 1 (skewed tastes), MVN-MXL is unable to recover the mode of the true heterogeneity distribution due to the restrictive symmetry property of the multivariate normal distribution. By contrast, 2-F-MON-MXL and DP-MON-MXL are more expressive: Both methods appear to identify two clusters, whereby the first cluster coincides with the mode of the true heterogeneity distribution, while the second cluster covers the tail of the true heterogeneity distribution. In scenario 2 (multi-modal, skewed tastes), MVN-MXL broadly envelops the true heterogeneity distribution. This is because the multivariate normal distribution cannot recover the true heterogeneity distribution due to its restrictive bell shape. By comparison, 2-F-MON-MXL and DP-MON-MXL are more flexible and perform visibly better at recovering the shape of the true heterogeneity distribution. 

\begin{table}[H]
\input{table_results_S1.tex}
\caption{Results for scenario 1 (skewed tastes)} \label{table_results_S1}
\end{table}

\begin{table}[H]
\input{table_results_S2.tex}
\caption{Results for scenario 2 (multi-modal, skewed tastes)} \label{table_results_S2}
\end{table}

\section{Case study} \label{S:case_study}

\subsection{Data}

Data for the case study are sourced from a stated choice study investigating preferences for features of shared automated vehicle (SAV) services in New York City \citep{bansal2018influence, liu2018framework}. The stated choice study was conducted in autumn 2017 and targeted the adult resident population of New York City, whereby individuals serving as drivers for MOD services were excluded from participation in the study. Valid responses were collected from a total of 1,507 individuals. A pivot-efficient approach was employed for the design of the choice tasks \citep[see][for details]{bansal2018influence}. Respondents were presented with seven choice tasks, each including three labelled alternatives, namely Uber (without ride-sharing/-splitting), UberPool (with ride-sharing/-splitting) and the current mode. The former two alternatives correspond to hypothetical SAV service options, which were deliberately named after existing brands of popular ride-sourcing services to increase the respondents' familiarity with the hypothetical SAV service options. The alternatives were characterised by six attributes, namely out-of-vehicle travel time (OVTT), in-vehicle travel time (IVTT), trip cost, parking cost, the powertrain of the vehicle (gas/petrol or electric), and the automation level of the vehicle (with or without driver). An example of a choice task is shown in Figure \ref{f_choice_task}. For the purpose of this case study, the data are randomly split into a training sample and a validation sample, and 20 of such random splits are taken. Each training sample contains observations from 1,207 individuals ($\approx 80\%$ of the total sample), while each validation sample includes one randomly selected choice task from each of the remaining 300 individuals ($\approx 20\%$ of the total sample).

\begin{figure}[h]
\footnotesize
\input{figure_choice_task.tex}
\caption{Example of a choice task \citep[reproduced from][]{liu2018framework}} \label{f_choice_task}
\end{figure}

\subsection{Methodology}

\subsubsection{Model specification and estimation} \label{subsubS:model_specification}

The primary objective of the case study is to infer distributions of willingness to pay (WTP) for features of MOD services. Hence, we specify utility in WTP space rather than in preference space so that WTP distributions can be directly estimated \citep[see][]{scarpa2008utility, train2005discrete}. To be specific, we employ a utility specification of the following form:
\begin{equation}
U_{ntj} =  \alpha_{n} d_{ntj} + \exp(\beta_{n}) \left ( - p_{ntj} + \boldsymbol{X}_{ntj} \boldsymbol{\gamma}_{n} \right) + \epsilon_{ntj}.
\end{equation}
Here, $n \in \{1, \ldots, N\}$ indexes decision-makers, $t \in \{1, \ldots, T\}$ indexes choice occasions and $j \in \{1, \ldots, J\}$ indexes alternatives. $\epsilon_{ntj}$ is a stochastic disturbance distributed $\text{Gumbel}(0,1)$. $d_{ntj}$ is a dummy variable indicating whether alternative $j$ is Uber or UberPool, $p_{ntj}$ is the total cost of alternative $j$, and $\boldsymbol{X}_{ntj}$ is a vector of other non-price attributes of alternative $j$, namely OVTT, IVTT, the interaction of IVTT and a dummy variable indicating whether alternative $j$ involves ride-sharing/-splitting (henceforth, ``shared $\times$ IVTT'') as well as whether the MOD vehicle is an electric vehicle (henceforth, ``electrification''), and whether the MOD vehicle is an automated vehicle (henceforth, ``automation''). $\alpha_{n}$, $\beta_{n}$,  $\boldsymbol{\gamma}_{n}$ are unknown, individual-specific utility parameters whose distributions we wish to infer. $\alpha_{n}$ is an alternative-specific constant, $\exp(\beta_{n})$ denotes the sensitivity to $- p_{ntj}$, and $\boldsymbol{\gamma}_{n}$ can be readily interpreted as WTP for the non-price attributes $\boldsymbol{X}_{ntj}$. For all modelling approaches, we assume that $\alpha_{n}$ and $\beta_{n}$ follow a normal distribution, whereby the correlation between $\alpha_{n}$ and $\beta_{n}$ is restricted to zero. $\beta_{n}$ enters the utility function exponentially to assure that sensitivities to $- p_{ntj}$ are strictly positive. For MVN-MXL, $\boldsymbol{\gamma}_{n}$ is assumed to come from a multivariate normal distribution with an unrestricted covariance matrix. For 2-F-MON- and DP-MON-MXL, $\boldsymbol{\gamma}_{n}$ follows a two-component mixture of normals distribution or, respectively, a Dirichlet process mixture of normals distribution. Regarding model estimation, the same practicalities as for the simulation study apply (see Section \ref{S:sim_study:data}).

\subsubsection{Model comparison and selection}

To compare the performance of the modelling approaches on the training and validation samples, we rely on two statistics, namely the log posterior predictive density \citep[LPPD;][]{gelman2014understanding} and the widely applicable information criterion \citep[WAIC;][]{gelman2014understanding, watanabe2013widely}. In contrast to other model selection criteria such as Akaike's information criterion (AIC) or the Bayesian information criterion (BIC), LPPD and WAIC are fully compatible with the Bayesian inference approach, as they account for posterior uncertainty. For a recent, in-depth treatment of model comparison and selection in a Bayesian context, we refer to \citet{gelman2014understanding}. 

In essence, LPPD is the point-wise log-likelihood integrated over the posterior draws of the model parameters. For the training sample, LPPD measures how well the trained model fits existing data. In this case, LPPD is given by 
\begin{equation}
\text{LPPD} = \sum_{n = 1}^{N} \sum_{t = 1}^{T_{n}} 
\ln \left (
\int P(y_{nt} \vert \boldsymbol{X}_{nt}, \boldsymbol{\beta}) p(\boldsymbol{\beta}_{n} \vert \boldsymbol{y}) d \boldsymbol{\beta}_{n}
\right ).
\end{equation}
For the validation sample, LPPD measures how well predictions generated from the trained model fit new data. In this case, LPPD is given by 
\begin{equation}
\text{LPPD} = \sum_{n = 1}^{N} \sum_{t = 1}^{T_{n}} 
\ln \left ( 
\hat{P} (y_{nt}^{*} \vert \boldsymbol{X}_{nt}^{*}, \boldsymbol{y})
\right ),
\end{equation}
where $\hat{P} (y_{nt}^{*} \vert \boldsymbol{X}_{nt}^{*}, \boldsymbol{y})$ are obtained either via expression \ref{e:ppd_mvn} (for MVN-MXL) or via expression \ref{e:ppd_mon} (for 2-F-MON- and DP-MON-MXL). In both cases, higher values of LPPD indicate better model fit.

Unlike LPPD, WAIC includes an adjustment for model complexity. To be specific, WAIC corrects LPPD for the effective number of model parameters $p_{\text{WAIC}}$, which is defined as
\begin{equation}
p_{\text{WAIC}} = \sum_{n = 1}^{N} \sum_{t = 1}^{T_{n}} 
\text{Var} \left \{ \ln P(y_{nt} \vert \boldsymbol{X}_{nt}, \boldsymbol{\beta}) \right \}
\end{equation}
and is computed by means of Monte Carlo integration over the posterior draws. Then, we have
\begin{equation}
\text{WAIC} = -2 \left ( \text{LPPD} - p_{\text{WAIC}}  \right ),
\end{equation}
where the pre-multiplication by $-2$ projects the information criterion onto the deviance scale so that WAIC can be interpreted in the same way as AIC and BIC, i.e. the model with the lowest WAIC value should be preferred over competing models. 

\subsection{Results}

First, we compare the in-sample fit and out-of-sample predictive ability of the considered models. Summary statistics for the estimated models are enumerated in Table \ref{table_modelComparison}. We report the means and standard errors of the considered performance metrics across 20 replications. For the training samples, LPPD indicates that DP-MON-MXL provides the best fit to the data, and WAIC suggests that the greater model complexity of the model is justified. As indicated by LPPD and WAIC, 2-F-MON-MXL is the second-best performing training samples. For the validation samples, the differences in the mean values of LPPD and WAIC across the considered models are not statistically significant (ANOVA for LPPD: $\text{df} = 2$, $F =  0.037$, $p = 0.964$; ANOVA for WAIC: $\text{df} = 2$, $F =  0.035$, $p = 0.965$). Estimation times are manageable for all methods. MVN-MXL is the fastest method, while DP-MON-MXL is the slowest method due to its higher complexity. 

\begin{landscape}
\begin{figure}[h]
\centering
\includegraphics[width = 1.4 \textwidth]{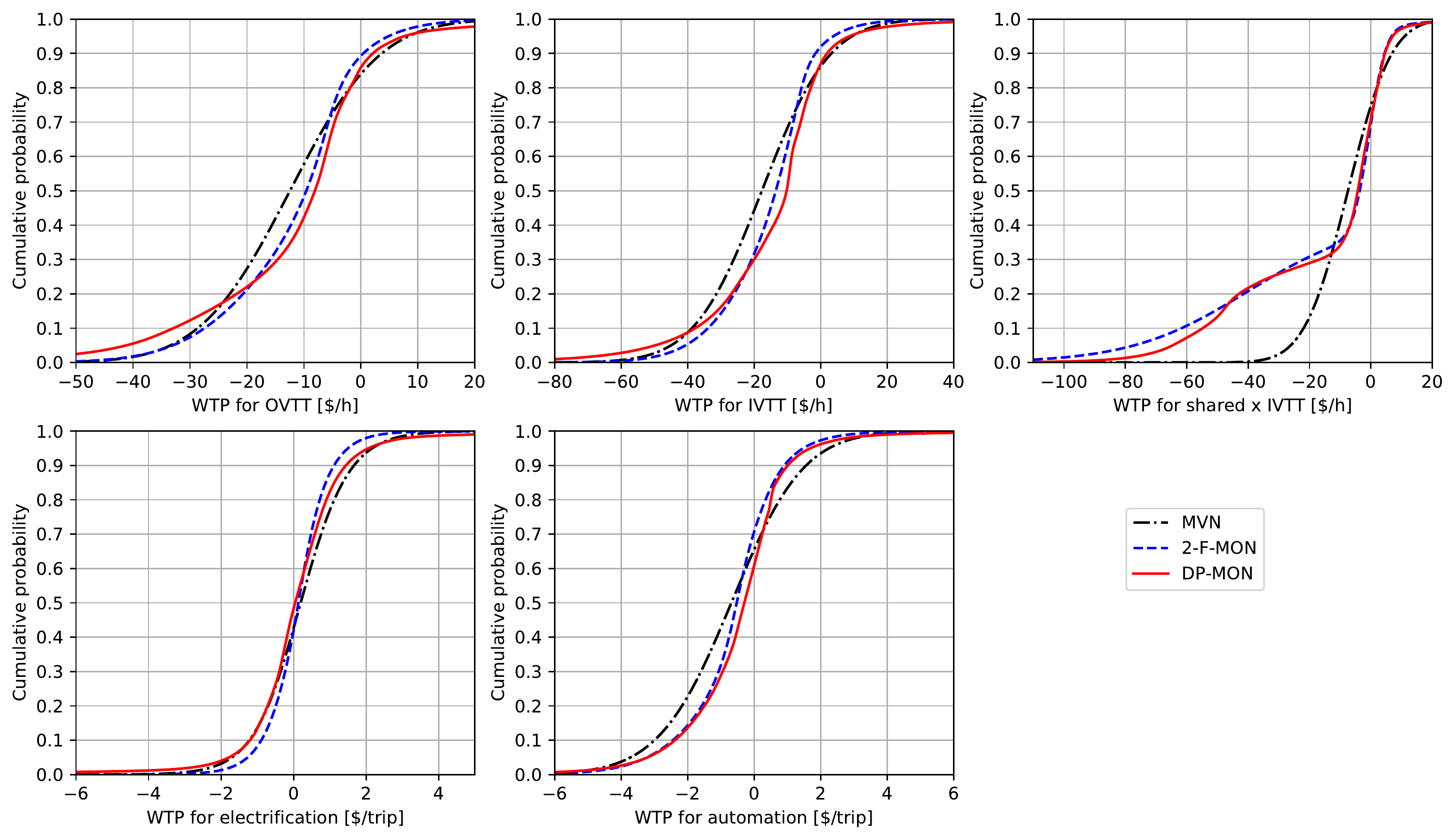}
\caption{Estimated posterior predictive cumulative distribution functions of willingness to pay (WTP) for features of mobility-on-demand services for one of the random splits} \label{f_cs_uni}
\end{figure}
\end{landscape}

Next, we contrast the heterogeneity representations produced by each of the considered models. Figure \ref{f_cs_uni} shows the estimated posterior predictive cumulative distribution functions (CDFs) of WTP for the non-price attributes for the considered models for one of the random splits. Moreover, Table \ref{table_cs_dist} enumerates summary statistics for the estimated heterogeneity distributions of all random parameters. Note that in accordance with the model specification outlined in Section \ref{subsubS:model_specification}, negative WTP indicates that a feature is undesirable, whereas positive WTP indicates that a feature is desirable. From Figure \ref{f_cs_uni}, it can be seen that the 2-F-MON and DP-MON mixing distributions afford greater distributional flexibility than the MVN mixing distribution, which is more restrictive due to its symmetric bell shape. Table \ref{table_cs_dist} shows that the considered models give closely similar estimates for the parametrically distributed random parameters $\alpha$ (alternative-specific constant) and $\beta$ (price sensitivity). 

In what follows, we discuss the estimated heterogeneity distributions of WTP for the non-price attributes in more detail. 
For OVTT, the DP-MON-MXL model suggests that the mean WTP is $-11.34$ \$/h. It can further be seen that the heterogeneity representations produced by the 2-F-MON and DP-MON mixing distributions have very similar shapes, whereas the MVN mixing distribution indicates lower mean and median WTP values. 
For IVTT, similar observations can be made. The 2-F-MON and DP-MON-MXL models yield closely overlapping representations of unobserved heterogeneity, while the MVN-MXL model suggests lower mean and median WTP values for the attribute in question. 
For the interaction of shared/split rides and IVTT, the parametric and semi-parametric representations of unobserved heterogeneity turn out to be drastically different from each other. The 2-F-MON- and DP-MON-MXL models suggest that there are two segments in the sample. One segment accounting for approximately two thirds of the sample is indifferent to ride-splitting or even desires to share a vehicle with others; the second segment accounting for the remaining third of the sample is averse to ride-splitting and demands comparatively large monetary compensations for sharing a vehicle with others. In fact, the DP-MON-MXL model suggests that approximately 30\% of the sample is willing to pay 20 \$/h or more to avoid ride-splitting. By contrast, the MVN-MXL model is unable to capture this extent of taste variation. Whereas the MVN-MXL model suggests a mean WTP of $-7.49$ \$/h, the 2-F-MON- and DP-MON-MXL models give considerably lower mean WTP estimates of $-17.21$ \$/h and $-15.28$ \$/h, respectively. 

Furthermore, the estimation results suggest that vehicle automation and powertrain electrification may be relatively unimportant to travellers. For vehicle electrification, the sample is roughly equally split between individuals who desire powertrain electrification and those who do not. However, WTP for powertrain electrification is generally small. All models give near-zero mean estimates of WTP. The DP-MON-MXL models estimates that approximately 20\% of the respondents are willing to pay at least 1 \$ per trip for travel by electric vehicle; another 20\% of the respondents demand a compensation of at least 1 \$ per trip to accept travelling by electric vehicle. For vehicle automation, all models suggest that the majority of the respondents do not desire to travel by automated vehicle. Approximately two thirds of the sample are not willing to pay excess fare to travel by automated vehicle; the DP-MON-MXL models suggests that less than 10\% of the respondents are willing to pay at least 1 \$ per trip to travel by automated vehicle. Consequently, vehicle automation and powertrain electrification may primarily provide indirect, rather than immediate benefits to users of SAV services by allowing for lower operating costs and increased operational efficiency \citep[see e.g.][]{burns2013transforming}. 

\begin{landscape}
\begin{table}[h]
\input{table_modelComparison.tex}
\caption{Model comparison} \label{table_modelComparison}
\end{table}
\end{landscape}

\begin{table}[h]
\small
\input{table_cs_dist.tex}
\caption{Estimated posterior predictive distributions of utility parameters for one of the random splits} \label{table_cs_dist}
\end{table}

\section{Conclusions}  \label{S:conclusion}

In this paper, we contrast the performance of different parametric and semi-parametric mixing distributions for mixed logit in a fully Bayesian setting. Specifically, we compare the multivariate normal (MVN), two-component finite mixture of normals (2-F-MON) and Dirichlet process mixture of normals (DP-MON) heterogeneity distributions, which are well-grounded in the hierarchical Bayesian modelling paradigm due to their conditionally-conjugate structure. The DP-MON mixing distribution promises to be particularly flexible in terms of the distributional shapes it can assume and does not require that its complexity is fixed prior to estimation unlike other semi-parametric approaches. However, the DP-MON mixing distribution is not widely used in travel demand analysis and its properties relative to simpler parametric and semi-parametric mixing distributions are not well understood. In this paper, we evaluate the performance of the considered mixing distributions on simulated data and on real data sourced from a stated choice study investigating preferences for shared automated vehicle (SAV) services in New York City. In each of the considered data settings, our analysis shows that the DP-MON mixing distribution offers superior fit to the data and performs at least as well as the competing methods at out-of-sample prediction.

The case study also provides several useful behavioural insights into the adoption of SAV services. The mixed logit models with flexible, semi-parametric representations of unobserved heterogeneity reveal that preferences for in-vehicle time by SAV with ride-splitting are strongly polarised. We estimate that there are two segments in the sample. While one segment is indifferent to ride-splitting or even desires it, a second segment accounting for the remaining third of the sample is willing to pay between 10 \$/h to 80 \$/h to avoid ride-splitting. This finding underscores that SAV services with and without ride-splitting are perceived as two different mobility options and suggests that there is potential for two distinct markets for SAV services with and without ride-splitting \citep[also see][]{krueger2016preferences, lavieri2019modeling, vij2018australian}. Moreover, the case study suggests that vehicle electrification and automation are relatively unimportant to travellers. The estimation results for the DP-MON-MXL model indicate that 80\% of the respondents are willing to pay between $-1.2$ \$ and $1.5$ \$ per trip for vehicle electrification; likewise, 80\% of the respondents are willing to pay between $-2.4$ \$ and $1.0$ \$ per trip for vehicle automation. This result suggests that vehicle automation and powertrain electrification may not be perceived as stand-alone competitive advantages by travellers. Rather, travellers may indirectly benefit from vehicle automation and electrification through increases in operational efficiency and lower operating costs \citep[see e.g.][]{burns2013transforming}.\footnote{Given the novelty of the considered vehicle technologies and mobility on demand services in general, we caveat the behavioural insights derived from the case study by acknowledging that the stated choice data are likely subject to hypothetical biases \citep{beck2016can, krueger2016preferences}. It remains to be shown whether the elicited preferences will persist once mobility on demand services with shared automated, electric vehicles become a reality. Nonetheless, we argue that stated choice data can provide useful insights into the relative importance of attributes.}

This study aimed at advancing the literature in two respects, namely the modelling of preferences for SAV services and the Bayesian estimation of mixed random utility models with flexible, semi-parametric mixing distributions. There are several directions in which future research can build on the current paper. First, our study showed that a non-negligible proportion of travellers is averse to sharing a vehicle with strangers. Yet, ride-splitting may be an effective means to increase capacity utilisation during peak travel times and to improve the environmental performance of SAV services \citep[see e.g.][]{fagnant2018dynamic}. Thus, future research may be directed at better understanding demand for pooled and non-pooled SAV services in order to inform policies targeted at encouraging the adoption of ride-splitting. One possible way to control for sources of unobserved preferences heterogeneity is to incorporate stochastic latent variables into the model formulation \citep[see][]{lavieri2019modeling}. Second, future research may focus on integrating disaggregate behavioural modelling approaches with flexible representations of unobserved heterogeneity with methods for the strategic and operational management of SAV service systems. For example, \citet{liu2018framework} demonstrate how supply side parameters of SAV service systems can be optimised taking into account disaggregate preferences; however, their framework does not accommodate unobserved preference heterogeneity. Finally, variational Bayes (VB) methods are emerging as fast and computationally-efficient alternative to MCMC methods for Bayesian estimation of mixed logit models \citep[see][]{bansal2019bayesian}. VB methods enable scaling disaggregate behavioural models to very large datasets \citep[see][]{blei2017variational}. However, existing VB methods for mixed logit are limited to models with multivariate normal mixing distributions. Consequently, future work may explore ways to enable VB for mixed logit models with more flexible, semi-parametric mixing distributions. 

\section*{Acknowledgements}

We would like to thank Prateek Bansal and Ricardo A. Daziano for sharing the stated choice data.
RK and THR acknowledge financial support from the Australian Research Council (DE170101346).
This research includes computations using the Linux computational cluster Katana supported by the Faculty of Science, UNSW Australia.

\section*{Author contribution statement}
RK: conception and design, model implementation, data preparation and analysis, manuscript writing and editing. \\
AV: conception and design, manuscript editing, supervision. \\
THR: conception and design, manuscript editing, supervision. \\

\newpage
\bibliographystyle{apalike}
\bibliography{bibliography.bib}

\newpage
\begin{appendices}

\section{Model estimation} \label{A:estimation}

\subsection{Mixed logit with a multivariate normal (MVN) mixing distribution}

Algorithm 1:
\begin{enumerate}
\item Update $\boldsymbol{\zeta}$ by sampling $\boldsymbol{\zeta} \sim \text{N}\left ( \boldsymbol{\mu}_{\boldsymbol{\zeta}}, \boldsymbol{\Sigma}_{\boldsymbol{\zeta}} \right )$, where
$\boldsymbol{\Sigma}_{\boldsymbol{\zeta}} = \left ( \boldsymbol{\Sigma}_{0} + N \boldsymbol{\Omega}^{-1} \right )^{-1}$ and
$\boldsymbol{\mu}_{\boldsymbol{\zeta}} = \boldsymbol{\Sigma}_{\boldsymbol{\zeta}} \Big ( \boldsymbol{\Sigma}_{0}^{-1} \boldsymbol{\mu}_{0} + \boldsymbol{\Omega}^{-1} \sum_{n=1}^{N} \boldsymbol{\beta}_{n} \Big )$

\item Update $a_{r}$ for all $r \in \{1, \ldots, R\}$ by sampling $a_{r} \sim \text{Gamma}\Big ( \frac{\nu + R}{2},  \frac{1}{A_{r}^{2}} + \nu \big( \boldsymbol{\Omega}^{-1} \big )_{rr} \Big )$.

\item Update $\boldsymbol{\Omega}$ by sampling $\boldsymbol{\Omega} \sim \text{IW} \left (\nu + N + K - 1, 2 \nu \text{diag}(\boldsymbol{a})  + \sum_{n=1}^{N} (\boldsymbol{\beta}_{n} -  \boldsymbol{\zeta}) (\boldsymbol{\beta}_{n} -  \boldsymbol{\zeta})^{\top} \right )$.

\item Update $\boldsymbol{\beta}_{n}$ for all $n \in \{1, \ldots, N\}$:

\begin{enumerate}
\item Propose $\tilde{\boldsymbol{\beta}}_{n} = \boldsymbol{\beta}_{n} + \sqrt{\rho} \text{chol}(\boldsymbol{\Omega}) \boldsymbol{\eta}$, where $\boldsymbol{\eta} \sim \text{N}(\boldsymbol{0},\boldsymbol{I}_{K})$.
\item Compute $r = 
\frac{P(\boldsymbol{y}_{n} \vert \boldsymbol{X}_{n}, \tilde{\boldsymbol{\beta}}_{n}) \phi( \tilde{\boldsymbol{\beta}}_{n} \vert \boldsymbol{\zeta},  \boldsymbol{\Omega})}
{P(\boldsymbol{y}_{n} \vert \boldsymbol{X}_{n}, \boldsymbol{\beta}_{n}) \phi( \boldsymbol{\beta}_{n} \vert \boldsymbol{\zeta},  \boldsymbol{\Omega})}$.
\item Draw $u \sim \text{Uniform}(0,1)$. If $r \leq u$, accept the proposal. If  $r > u$, reject the proposal. 
\end{enumerate}

\end{enumerate}
$\rho$ is a step size, which needs to be tuned. Here, we employ the same tuning mechanism as \citet{train2009discrete}: $\rho$ is set to an initial value of 0.1 and after each iteration, $\rho$ is decreased by 0.001, if the average acceptance rate across all decision-makers is less than 0.3; $\rho$ is increased by 0.001, if the average acceptance rate across all decision-makers is more than 0.3. 

\subsection{Mixed logit with a finite mixture of normals (F-MON) mixing distribution}

Algorithm 2:
\begin{enumerate}
\item Update $\boldsymbol{\zeta}_{k}$ for all $k \in \{1, \ldots, K\}$ by sampling $\boldsymbol{\zeta}_{k} \sim \text{N}\left ( \boldsymbol{\mu}_{\boldsymbol{\zeta}_{k}}, \boldsymbol{\Sigma}_{\boldsymbol{\zeta}_{k}} \right )$, where
$\boldsymbol{\Sigma}_{\boldsymbol{\zeta}_{k}} = \left ( \boldsymbol{\Sigma}_{0} + c_{k} \boldsymbol{\Omega}^{-1} \right )^{-1}$ and
$\boldsymbol{\mu}_{\boldsymbol{\zeta}_{k}} = \boldsymbol{\Sigma}_{\boldsymbol{\zeta}_{k}} \left ( \boldsymbol{\Sigma}_{0}^{-1} \boldsymbol{\mu}_{0} + \boldsymbol{\Omega}^{-1} \sum_{n : q_{n} = k} \boldsymbol{\beta}_{n} \right )$

\item Update $a_{kr}$ for all $k \in \{1, \ldots, K\}$ and $r \in \{1, \ldots, R\}$ by sampling $a_{kr} \sim \text{Gamma}\Big ( \frac{\nu + R}{2},  \frac{1}{A_{r}^{2}} + \nu \big ( \boldsymbol{\Omega}_{k}^{-1} \big )_{rr} \Big )$.

\item Update $\boldsymbol{\Omega}_{k}$ for all $k \in \{1, \ldots, K\}$ by sampling $\boldsymbol{\Omega}_{k} \sim \text{IW} (\nu + c_{k} + R - 1, 2 \nu \text{diag}(\boldsymbol{a}_{k})  + \sum_{n : q_{n} = k} (\boldsymbol{\beta}_{n} -  \boldsymbol{\zeta}_{k}) (\boldsymbol{\beta}_{n} -  \boldsymbol{\zeta}_{k})^{\top} )$.

\item Update $\boldsymbol{\pi}$ by sampling $\boldsymbol{\pi} \sim \text{Dirichlet}(\boldsymbol{\bar{\alpha}})$, where $\bar{\alpha}_{k} = \alpha + c_{k}$.

\item Update $q_{n}$ for all $n \in \{1, \ldots, N\}$ by sampling $q_{n} \sim \text{Categorical}(\boldsymbol{p})$, where $p_{k} = \frac{ \pi_{k} \phi( \boldsymbol{\beta}_{n} \vert \boldsymbol{\zeta}_{k},  \boldsymbol{\Omega}_{k})}{\sum_{k' = 1}^{K} \pi_{k'} \phi( \boldsymbol{\beta}_{n} \vert \boldsymbol{\zeta}_{k'},  \boldsymbol{\Omega}_{k'})}$

\item Update $\boldsymbol{\beta}_{n}$ for all $n \in \{1, \ldots, N\}$:

\begin{enumerate}
\item Propose $\tilde{\boldsymbol{\beta}}_{n} = \boldsymbol{\beta}_{n} + \sqrt{\rho} \text{chol}(\boldsymbol{\Omega}) \boldsymbol{\eta}$, where $\boldsymbol{\eta} \sim \text{N}(\boldsymbol{0},\boldsymbol{I}_{K})$.
\item Compute $r = 
\frac{P(\boldsymbol{y}_{n} \vert \boldsymbol{X}_{n}, \tilde{\boldsymbol{\beta}}_{n}) \phi( \tilde{\boldsymbol{\beta}}_{n} \vert \boldsymbol{\zeta}_{k},  \boldsymbol{\Omega}_{k})}
{P(\boldsymbol{y}_{n} \vert \boldsymbol{X}_{n}, \boldsymbol{\beta}_{n}) \phi( \boldsymbol{\beta}_{n} \vert \boldsymbol{\zeta}_{k},  \boldsymbol{\Omega}_{k})}$.
\item Draw $u \sim \text{Uniform}(0,1)$. If $r \leq u$, accept the proposal. If  $r > u$, reject the proposal. 
\end{enumerate}

\end{enumerate}
The step size $\rho$ is tuned in the same way as for Algorithm 1. 

\subsection{Mixed logit with a Dirichlet process mixture of normals (DP-MON) mixing distribution}

The MCMC algorithm for mixed logit with DP-MON mixing distribution is the same as Algorithm 2, the only difference being that step 4 is replaced with the following steps:
\begin{enumerate}
\item Update $\alpha$ by sampling $\alpha \sim \text{Gamma} \left (2 + K - 1, 2 - \sum_{k=1}^{K - 1} \ln (1 - \eta_{k}) \right )$.

\item Update $\eta_{k}$ for all $k \in \{1, \ldots, K - 1\}$ by sampling $\eta_{k} \sim \text{Beta}(1 + c_{k}, \alpha + \sum_{j=k+1}^{K} c_{j})$, set $\eta_{K} = 1$, and calculate $\pi_{k} = \eta_{k} \prod_{l=1}^{k-1}(1 - \eta_{l})$ for all $k \in \{1, \ldots, K\}$.
\end{enumerate}

\end{appendices}

\end{document}

%% file: table_results_S1.tex
\centering
\ra{1.2}
\begin{tabular}{@{} l 
S[table-format=4.1] S[table-format=4.1]  c 
S[table-format=1.4] S[table-format=1.4]  @{}} 
\toprule

& 
\multicolumn{2}{c}{\textbf{Estimation time [s]}} & &
\multicolumn{2}{c}{\textbf{TVD [\%]}} \\
\cmidrule{2-3} \cmidrule{5-6}

& 
\textbf{Mean} & \textbf{Std. err.} & &
\textbf{Mean} & \textbf{Std. err.} \\
\midrule

$N = 1000$; $T = 8$ \\
\quad MVN &  311.6 &  2.8 & &0.3774 & 0.0133 \\ 
\quad 2-F-MON &  838.6 &  7.8 & &0.2411 & 0.0136 \\ 
\quad DP-MON & 6376.1 & 52.9 & &0.2056 & 0.0146 \\ 
$N = 1000$; $T = 16$ \\
\quad MVN &  551.9 &  6.6 & &0.3845 & 0.0117 \\ 
\quad 2-F-MON & 1050.2 & 11.6 & &0.2625 & 0.0136 \\ 
\quad DP-MON & 6617.0 & 33.1 & &0.1969 & 0.0162 \\ 

\midrule
\multicolumn{6}{l}{
\begin{minipage}[t]{0.6 \textwidth}
\footnotesize
Note: 
TVD = total variation distance between true and predicted choice probabilities for a validation sample. The reported means and standard errors are based on 20 replications.
\end{minipage}} \\
\bottomrule
\end{tabular}

%% file: table_results_S2.tex
\centering
\ra{1.2}
\begin{tabular}{@{} l 
S[table-format=4.1] S[table-format=4.1]  c 
S[table-format=1.4] S[table-format=1.4]  @{}} 
\toprule

& 
\multicolumn{2}{c}{\textbf{Estimation time [s]}} & &
\multicolumn{2}{c}{\textbf{TVD [\%]}} \\
\cmidrule{2-3} \cmidrule{5-6}

& 
\textbf{Mean} & \textbf{Std. err.} & &
\textbf{Mean} & \textbf{Std. err.} \\
\midrule

$N = 1000$; $T = 8$ \\
\quad MVN &  306.0 &  5.5 & &1.0155 & 0.0219 \\ 
\quad 2-F-MON &  836.9 &  4.9 & &0.4584 & 0.0239 \\ 
\quad DP-MON & 6434.6 & 52.3 & &0.3615 & 0.0190 \\ 
$N = 1000$; $T = 16$ \\
\quad MVN &  560.6 &  9.1 & &1.1084 & 0.0302 \\ 
\quad 2-F-MON & 1055.1 & 11.0 & &0.5368 & 0.0316 \\ 
\quad DP-MON & 6707.8 & 21.7 & &0.3957 & 0.0391 \\ 

\midrule
\multicolumn{6}{l}{
\begin{minipage}[t]{0.6 \textwidth}
\footnotesize
Note: 
See Table \ref{table_results_S1} for an explanation of the table headers.

\end{minipage}} \\
\bottomrule
\end{tabular}

%% file: figure_choice_task.tex
\centering
\ra{1.2}
\begin{tabular}{@{} l c c c@{}} 
& \textbf{Uber (with ride-sharing)} & \textbf{UberPool (without ride-sharing)} & \textbf{Current mode: car} \\
\toprule
Walking and waiting time & 6 min & 9 min & 12 min \\
In-vehicle travel time & 38 min & 50 min & 48 min \\
Trip cost (excl. parking cost) & \$11 & \$8 & \$6 \\
Parking cost & -- & -- & \$6 \\
Powertrain & Electric & Gas & Gas \\
Automation & Service with driver & Automated (no driver) & -- \\
\bottomrule
\end{tabular}

%% file: table_modelComparison.tex
\centering
\ra{1.2}
\begin{tabular}{@{} l 
S[table-format=4.1] S[table-format=2.1]  c 
S[table-format=4.2] S[table-format=1.2]  c 
S[table-format=5.2] S[table-format=2.2]  c 
S[table-format=3.2] S[table-format=1.2]  c 
S[table-format=3.2] S[table-format=1.2]  @{}} 
\toprule
& & & & \multicolumn{5}{c}{\textbf{Training}} & & \multicolumn{5}{c}{\textbf{Validation}} \\
\cmidrule{5-9} \cmidrule{11-15} 
& 
\multicolumn{2}{c}{\textbf{Estimation time [s]}} & &
\multicolumn{2}{c}{\textbf{LPPD}} & &
\multicolumn{2}{c}{\textbf{WAIC}} & &
\multicolumn{2}{c}{\textbf{LPPD}} & &
\multicolumn{2}{c}{\textbf{WAIC}} \\
\cmidrule{2-3} \cmidrule{5-6} \cmidrule{8-9} \cmidrule{11-12} \cmidrule{14-15}

& 
\textbf{Mean} & \textbf{Std. err.} & &
\textbf{Mean} & \textbf{Std. err.} & &
\textbf{Mean} & \textbf{Std. err.} & &
\textbf{Mean} & \textbf{Std. err.} & &
\textbf{Mean} & \textbf{Std. err.} \\
\midrule

MVN &  442.3 &   8.1 & &-3534.95 & 13.32 & &9862.77 & 24.63 & &-274.66 & 2.71 & &550.53 & 5.44 \\ 
2-F-MON & 1623.1 &  44.0 & &-3512.56 & 13.76 & &9682.71 & 26.94 & &-273.73 & 2.77 & &548.70 & 5.57 \\ 
DP-MON & 7594.6 & 181.8 & &-3431.76 & 12.81 & &9588.38 & 25.64 & &-273.75 & 2.75 & &548.78 & 5.52 \\ 

\midrule
\multicolumn{15}{l}{
\begin{minipage}[t]{1 \textwidth}
\footnotesize
Note: 
The reported means and standard errors are based on 20 replications.
\end{minipage}} \\
\bottomrule
\end{tabular}

%% file: table_cs_dist.tex
\centering
\ra{1.2}
\begin{tabular}{@{} l S[table-format=2.2] S[table-format=2.2] S[table-format=2.2]  S[table-format=2.2] S[table-format=2.2] c S[table-format=2.2] S[table-format=2.2] @{}}
\toprule
& \multicolumn{5}{c}{\textbf{Percentiles}} & \phantom{a} & \multicolumn{2}{c}{\textbf{Mean}}  \\ 
\cmidrule{2-6} \cmidrule{8-9} 
& 
\multicolumn{1}{c}{\textbf{10\textsuperscript{th}}}  &
\multicolumn{1}{c}{\textbf{25\textsuperscript{th}}}  & 
\multicolumn{1}{c}{\textbf{50\textsuperscript{th}}}  & 
\multicolumn{1}{c}{\textbf{75\textsuperscript{th}}}  & 
\multicolumn{1}{c}{\textbf{90\textsuperscript{th}}}  & & 
\multicolumn{1}{c}{\textbf{Est.}}  & 
\multicolumn{1}{c}{\textbf{Std.}}  \\
\midrule

$\alpha$\\
\quad MVN	&	-4.50	&	-2.74	&	-0.83	&	1.08	&	2.84	& &	-0.83	&	0.14	\\
\quad 2-F-MON	&	-4.25	&	-2.59	&	-0.73	&	1.13	&	2.84	& &	-0.72	&	0.13	\\
\quad DP-MON	&	-4.25	&	-2.59	&	-0.78	&	1.08	&	2.74	& &	-0.76	&	0.13	\\
$\beta$\\
\quad MVN	&	-3.64	&	-2.64	&	-1.55	&	-0.48	&	0.53	& &	-1.55	&	0.09	\\
\quad 2-F-MON	&	-3.84	&	-2.74	&	-1.53	&	-0.38	&	0.73	& &	-1.56	&	0.10	\\
\quad DP-MON	&	-3.84	&	-2.74	&	-1.48	&	-0.28	&	0.83	& &	-1.49	&	0.10	\\
WTP for OVTT [\$/h] \\
\quad MVN	&	-28.54	&	-20.80	&	-12.48	&	-3.92	&	3.47	& &	-12.48	&	1.21	\\
\quad 2-F-MON	&	-27.49	&	-17.99	&	-9.55	&	-4.62	&	0.30	& &	-11.68	&	1.10	\\
\quad DP-MON	&	-32.76	&	-17.64	&	-8.14	&	-3.57	&	2.06	& &	-11.34	&	1.21	\\
WTP for IVTT [\$/h] \\
\quad MVN	&	-38.39	&	-28.74	&	-17.71	&	-7.04	&	3.22	& &	-17.71	&	1.68	\\
\quad 2-F-MON	&	-33.57	&	-23.32	&	-13.67	&	-7.04	&	-1.61	& &	-15.57	&	1.64	\\
\quad DP-MON	&	-37.79	&	-23.32	&	-10.05	&	-4.62	&	2.01	& &	-14.61	&	1.35	\\
WTP for shared $\times$ IVTT [\$/h] \\
\quad MVN	&	-21.81	&	-15.28	&	-7.49	&	-0.25	&	6.93	& &	-7.49	&	0.82	\\
\quad 2-F-MON	&	-61.66	&	-32.26	&	-3.52	&	1.06	&	4.32	& &	-17.21	&	2.43	\\
\quad DP-MON	&	-54.47	&	-32.26	&	-4.17	&	1.06	&	4.32	& &	-15.28	&	1.89	\\
WTP for electrification [\$/trip] \\
\quad MVN	&	-1.25	&	-0.53	&	0.21	&	0.91	&	1.68	& &	0.21	&	0.19	\\
\quad 2-F-MON	&	-0.91	&	-0.36	&	0.14	&	0.58	&	1.13	& &	0.12	&	0.16	\\
\quad DP-MON	&	-1.19	&	-0.53	&	0.03	&	0.74	&	1.46	& &	0.09	&	0.17	\\
WTP for automation [\$/trip] \\
\quad MVN	&	-2.98	&	-1.84	&	-0.70	&	0.45	&	1.54	& &	-0.70	&	0.23	\\
\quad 2-F-MON	&	-2.44	&	-1.30	&	-0.51	&	0.15	&	0.93	& &	-0.64	&	0.24	\\
\quad DP-MON	&	-2.38	&	-1.18	&	-0.27	&	0.39	&	0.99	& &	-0.47	&	0.16	\\

\midrule
\multicolumn{9}{l}{{
\footnotesize 
Note:
$\alpha$: (alternative-specific constant); 
$\beta$: (normal random variable underlying sensitivity to price)}} \\
\bottomrule
\end{tabular}